\pdfoutput=1
\newcommand*{\ATLASLATEXPATH}{}
\documentclass[cernpreprint,texlive=2011,txfonts,UKenglish]{\ATLASLATEXPATH atlasdoc}

\usepackage{\ATLASLATEXPATH atlaspackage}
\usepackage{\ATLASLATEXPATH atlasbiblatex}

\usepackage{\ATLASLATEXPATH atlascontribute}

\usepackage[BSM,other,process,unit=true]{\ATLASLATEXPATH atlasphysics}

\usepackage{multirow}

\graphicspath{{.}}

\usepackage{DarkMatter-defs}

\AtlasTitle{Dark matter interpretations of ATLAS searches for the electroweak production of supersymmetric particles in $\sqrt{s} = 8\,$TeV proton--proton collisions}

\author{The ATLAS Collaboration}

\AtlasRefCode{SUSY-2015-12}

\AtlasJournal{JHEP}

 \PreprintIdNumber{CERN-PH-EP-2016-165}

\AtlasAbstract{%
A selection of searches by the ATLAS experiment at the LHC for the electroweak production of SUSY particles are used to study their impact on the constraints on dark matter candidates.
The searches use $20\,{\rm fb}^{-1}$ of proton--proton collision data at $\sqrt{s}=8\,$TeV.
A likelihood-driven scan of a five-dimensional effective model focusing on the gaugino--higgsino and Higgs sector of the phenomenological minimal supersymmetric Standard Model is performed.
This scan uses data from direct dark matter detection experiments, the relic dark matter density and precision flavour physics results.
Further constraints from the ATLAS Higgs mass measurement and SUSY searches at LEP are also applied.
A subset of models selected from this scan are used to assess the impact of the selected ATLAS searches in this five-dimensional parameter space.
These ATLAS searches substantially impact those models for which the mass $m(\tilde{\chi}^0_1)$ of the lightest neutralino is less than 65 GeV, excluding 86\% of such models.
The searches have limited impact on models with larger $m(\tilde{\chi}^0_1)$ due to either heavy electroweakinos or compressed mass spectra where the mass splittings between the produced particles and the lightest supersymmetric particle is small.
}

\AtlasCoverSupportingNote{Support note}{https://cds.cern.ch/record/2109714}
\AtlasCoverCommentsDeadline{14 July 2016}

\AtlasCoverAnalysisTeam{Gianfranco~Bertone, Christopher~Bock, Trygve~Buanes, Mike~Flowerdew, Zara~Grout, Lukas~Heinrich, Sebastian~Liem, Alexander~Mann, Alaettin~Serhan~Mete, Brian~Petersen, Christina~Jane~Potter, Roberto~Ruiz~de~Austri, Itzebelt~Santoyo~Castillo, Jeroen~Schouwenberg, Roberto~Trotta, Sarah~Williams}

\AtlasCoverEdBoardMember{Alan~Barr~(chair)}
\AtlasCoverEdBoardMember{Geraldine~Conti}
\AtlasCoverEdBoardMember{Heidi~Sandaker}

\AtlasCoverEgroupEditors{atlas-susy-2015-12-editors@cern.ch}

\AtlasCoverEgroupEdBoard{atlas-susy-2015-12-editorial-board@cern.ch}

\begin{document}
\maketitle

\tableofcontents
\newpage

\section{Introduction}

Supersymmetry, or SUSY~\cite{Golfand:1971iw,Volkov:1973ix,Wess:1974tw,Wess:1974jb,Ferrara:1974pu,Salam:1974ig}, is a popular candidate for physics beyond the Standard Model.
It provides an elegant solution to the hierarchy problem, which, in the Standard Model, demands high levels of fine tuning to counteract large quantum corrections to the mass of the Higgs boson~\cite{Weinberg:1975gm,Gildener:1976ai,Weinberg:1979bn,Susskind:1978ms}.
$R$-parity-conserving supersymmetric models can also provide a candidate for dark matter, in the form of the lightest supersymmetric particle (LSP)~\cite{Goldberg:1983nd,Ellis:1983ew}.

The ATLAS and CMS experiments performed a large number of searches for SUSY during Run-1 of the LHC and, in the absence of a significant excess in any channel, exclusion limits on the masses of SUSY particles (sparticles) were calculated in numerous scenarios, usually in the context of the minimal supersymmetric Standard Model (MSSM)~\cite{Fayet:1976et,Fayet:1977yc}.
These scenarios include ``high-scale'' SUSY models such as mSUGRA~\cite{Chamseddine:1982jx,Barbieri:1982eh,Kane:1993tda} or GMSB~\cite{Dine:1981gu,AlvarezGaume:1981wy,Nappi:1982hm}, both of which specify a particular SUSY-breaking mechanism.
Most searches also considered specific ``simplified models'', which attempt to capture the behaviour of a small number of kinematically accessible SUSY particles, often through considering one particular SUSY production process with a fixed decay chain.

Although the high-scale and simplified model exclusions provide an easily interpretable picture of the sensitivity of analyses to specific areas of parameter space, they are far from a full exploration of the MSSM, which contains about 120 free parameters.
The number of parameters is reduced if the phenomenological MSSM (pMSSM) is considered instead.
It is based on the most general CP-conserving MSSM, with $R$-parity conservation, and minimal flavour violation~\cite{Djouadi:1998di,Berger:2008cq}.
In addition, the first two generations of sfermions are required to be degenerate and have negligible Yukawa couplings.
This leaves 19 independent weak-scale parameters to be considered:
ten sfermion masses (five for the degenerate first two generations and five for the third generation),
three trilinear couplings $A_{\tau,t,b}$ which give the couplings between the Higgs field and the third-generation sfermions,
the bino, wino and gluino mass parameters $M_{1,2,3}$,
the higgsino mass parameter $\mu$,
the ratio of the vacuum expectation values of the Higgs fields $\tan\beta$,
and the mass of the pseudoscalar Higgs boson $m_{A}$.

The model considered here, henceforth referred to as EWKH, is described by only five parameters: $M_1$, $M_2$, $\mu$, and $\tan\beta$ to define the gaugino--higgsino sector, and $m_A$ to define the Higgs sector.
Both sectors are defined at tree level.
The coloured SUSY particles and sleptons are assumed to be heavy such that they do not impact the phenomenology.
This model is well motivated from a dark matter perspective since the dark matter candidate of the MSSM is the lightest neutralino whose properties are fully specified by these five parameters.
These parameters therefore also determine the relic density of the neutralino for much of the pMSSM parameter space, i.e. if coannihilations with slepton, squarks and gluinos are neglected.

An interpretation of the Run-1 SUSY searches in pMSSM models may be found in the literature (for instance Refs. \cite{Strege:2014ija,deVries:2015hva,Khachatryan:2016nvf}).
In particular, ATLAS has previously performed a study using about $300\,000$ pMSSM model points~\cite{SUSY-2014-08}.
In that work, all 19 of the pMSSM parameters were varied and the strongest direct constraints on sparticle production were obtained in searches for squarks and gluinos.
In this article, attention is restricted to a five-dimensional (5D) sub-space of the pMSSM in order to assess the impact of the ATLAS Run-1 searches (using $20\,\ifb$ of data at $\sqrt{s}=8\TeV$) specifically on the electroweak production of SUSY particles, and the corresponding constraints on dark matter. This provides a study complementary to that in Ref.~\cite{SUSY-2014-08} by decoupling strong-interaction production processes from the phenomenology, and thus allows more extensive exploration of the regions of parameter space relevant to electroweak production.
The scanning strategy used to select models is also different to Ref.~\cite{SUSY-2014-08},
where models were sampled from uniform distributions in the pMSSM parameters, and then required to satisfy a variety of experimental constraints.
In this study, an ``initial likelihood scan'' is performed to select models, using constraints from direct dark matter searches, precision electroweak measurements, flavour-physics results,
previous collider searches, and the ATLAS Higgs boson mass measurement.

The impact of the ATLAS searches in different regions of parameter space is established by considering the number of models selected by the initial likelihood scan that are excluded by the ATLAS electroweak SUSY searches.
Exclusion limits are calculated using the \cls\ technique~\cite{CLS}.
Both particle-level\footnote{Particle-level information constructs observables using the stable particles from MC generators, which account for the majority of interactions with the detector material~\cite{ATL-PHYS-PUB-2015-013}.} and reconstruction-level information is used to calculate the \cls\ values (see Section~\ref{sec:signalLH}), where the reconstruction-level information makes use of the ATLAS detector simulation, data-driven background estimations, and systematic uncertainties and their correlations. The \cls\ calculations invoke the simplifying assumption that the reconstruction of events selected at particle level can be parameterised using an average efficiency factor that does not depend on the details of the SUSY model.
The reconstruction-level information can then be used to directly map particle-level results to \cls\ values. This ``calibration procedure'' significantly reduces the computational load of the analysis and accounts, on average, for the acceptance and efficiency across the ensemble of models.

\section{ATLAS searches}
\label{sec:ATLASsearches}

Four ATLAS Run-1 SUSY searches that target electroweak SUSY production are considered, as listed in Table~\ref{tab:susySearches}.
Their combined impact on simplified models of electroweak sparticle production, as well as selected pMSSM and high-scale models, is summarised in Ref.~\cite{Aad:2015eda}.

The 2$\ell$ analysis~\cite{SUSY-2013-11} targets \slepton-pair production and \chinoonep{}\chinoonem production (where \chinoonepm decays via sleptons) with three signal regions, looking for an excess of events with $e^+e^-$, $\mu^+\mu^-$ or $e^\pm\mu^\mp$ and high stransverse mass (\mttwo)~\cite{Lester:1999tx,Barr:2003rg}.
Three additional signal regions target the more difficult scenario of \chinoonep{}\chinoonem production where the charginos decay via \Wboson bosons.
Finally, a seventh signal region requiring an opposite-sign light-lepton pair ($e^+e^-$, $\mu^+\mu^-$) with an invariant mass consistent with a \Zboson boson and an additional pair of jets is used to target \chinoonepm{}\ninotwo production where the chargino decays via a \Wboson boson and the neutralino decays via a \Zboson boson.
The 2$\tau$ analysis~\cite{SUSY-2013-14} uses four signal regions to search for \stau-pair, \chinoonep{}\chinoonem and \chinoonepm{}\ninotwo production, where the charginos and neutralinos decay via third-generation sleptons.
Events with a pair of opposite-sign hadronically decaying $\tau$-leptons (\tauhad) and large \mttwo\ are selected for the search. The 3$\ell$ analysis~\cite{SUSY-2013-12} searches for weakly interacting SUSY particles in events with three light leptons ($e/\mu$), two light leptons and one \tauhad, or one light lepton and two \tauhad.
Twenty-four signal regions are defined to target \chinoonepm{}\ninotwo production, where charginos and neutralinos decay via sleptons, staus, or the SM bosons \Wboson, \Zboson and $h$.
The 4$\ell$ analysis~\cite{SUSY-2013-13} searches for higgsino-like $\ninotwo\ninothree$ production, where the neutralinos decay via sleptons, staus or \Zboson bosons.
Nine signal regions are used to select events with large missing transverse momentum (whose magnitude is denoted as \met) and four light leptons, three light leptons and one \tauhad, or two light leptons and two \tauhad.

Although this article is restricted to these four analyses, other SUSY searches could provide sensitivity in some regions of the parameter space considered in this article.
For example, the ATLAS disappearing-track analysis~\cite{SUSY-2013-01} targets direct long-lived charginos with proper lifetimes $\mathcal{O}(1\,{\rm ns})$ so it could have sensitivity to compressed models where the mass difference of the lightest chargino and the LSP is much less than $1\GeV$. Consideration of this analysis is beyond the scope of this article.
Furthermore, the ATLAS monojet search~\cite{EXOT-2011-20} targets pair-produced dark matter particles but makes no assumption of an underlying supersymmetric theory. These results do not yet have sensitivity to direct electroweak SUSY production so is not considered further in this analysis.
\begin{table}[h]
\centering
\begin{tabular}{ll}
\toprule
Analysis & Target production processes \\
\midrule
2$\ell$~\cite{SUSY-2013-11}   & $\chinoonep\chinoonem$, $\chinoonepm\ninotwo$, $\slepton\slepton$      \\
2$\tau$~\cite{SUSY-2013-14}   & $\chinoonep\chinoonem$, $\chinoonepm\ninotwo$, $\stau\stau$      \\
3$\ell$~\cite{SUSY-2013-12}   & $\chinoonepm\ninotwo$     \\
4$\ell$~\cite{SUSY-2013-13}   & $\ninotwo\ninothree$     \\
\bottomrule
\end{tabular}
\caption{ATLAS electroweak SUSY searches re-interpreted in the pMSSM for this article.\label{tab:susySearches}}
\end{table}

\section{Theoretical framework}
\label{sec:theoreticalFramework}
The theoretical SUSY framework used in this article is an effective model of the electroweak gauginos, higgsinos and the Higgs sector of the MSSM, collectively labelled EWKH.
The model is described by five parameters, where four of them define the gaugino--higgsino sector at tree level ($M_1$, $M_2$, $\mu$, and $\tan \beta$), and $m_A$ is added to define the Higgs sector at tree level.
The other soft sparticle masses are large to ensure that the sfermions and gluinos are decoupled from the effective theory, while the trilinear couplings are not constrained. The specific values used are $5\TeV$ for the sfermion soft-masses, $4\TeV$ for the gluino mass and $0.1\TeV$ for the trilinear couplings.

When scanning in this framework, a Bayesian prior distribution for these parameters is used as a device to concentrate the parameter scan in certain regions of parameter space.
Two different prior distributions are adopted: ``flat priors'' are uniform in all model parameters, while ``log priors'' are uniform in the logarithm of all model parameters, except for $\tan \beta$, for which a uniform prior is used for both sets.
Flat priors tend to concentrate sampling towards large values of the parameters (as most of volume of the prior lies there), while log priors concentrate their scan in the lower mass ($\ll 1\TeV$) region (since this metric gives every decade in the parameter values the same a priori probability).
The posterior samples resulting from the flat and log prior scans are then merged to achieve a reliable mapping of the (prior-independent) profile likelihood function, as advocated in Ref.~\cite{Feroz:2011bj}.
Table~\ref{tab:EWKH_priors}  displays both of the priors used and their ranges.\footnote{For parameters that span both negative and positive numbers the log prior is actually a piecewise function in order to be invertible. The log parameter $\theta_i'$ is mapped onto the linear, physical, parameter $\theta_i$ as follows: if $|\theta_i'| \geq \log_{10} \mathrm{e}$ then $\theta_i = \mathrm{sign}(\theta_i') 10^{|\theta_i'|}$, otherwise $\theta_i = \theta_i'\mathrm{e}/ \log_{10} \mathrm{e}$.} The specific ranges are chosen because they contain the interesting dark matter phenomenology.

\begin{table}
\centering
\begin{tabular}{l l | l l}
\toprule
\multicolumn{2}{c}{Flat priors} & \multicolumn{2}{c}{Log priors} \\
\midrule
$M_1$ [\TeV] & $(-4, 4)$ & $\mathrm{sign} (M_1) \log_{10} |M_1|/$\GeV & $(-3.6, 3.6)$ \\
$M_2 $ [\TeV] & $(0.01, 4)$ & $\log_{10} M_2/$\GeV & $(1,3.6)$ \\
$\mu $ [\TeV] & $(-4,4)$ & $\mathrm{sign} (\mu) \log_{10} |\mu|/$\GeV & $(-3.6, 3.6)$ \\
$m_A $ [\TeV] & $(0.01, 4)$ & $\log_{10} m_A/$\GeV & $(1, 3.6)$ \\
$\tan\beta$ & $(2, 62)$ & $\tan\beta$ & $(2, 62)$  \\
\bottomrule
\end{tabular}
\caption{EWKH parameters used in the initial likelihood scan and the prior ranges for the two prior choices adopted.
``Flat priors'' are uniform in the parameter itself within the indicated ranges, while ``log priors'' are uniform in the logarithm of the parameter within the indicated ranges.
The physical ranges for both priors are identical for both the ``flat'' and ``log'' priors.
\label{tab:EWKH_priors} }
\end{table}

The profile likelihood maps obtained from merging the samples gathered with both priors explore in detail both the low-mass and the high-mass regions, for a more thorough scanning of the entire parameter space.

\subsection{Scanning strategy}
\label{sec:scanning}
A Bayesian approach is adopted for sampling the EWKH parameter space, and the sensitivity of the ATLAS SUSY electroweak analyses is calculated for the resulting posterior samples.
This ``initial likelihood scan'' is driven by the likelihood defined in Section \ref{sec:nonsusydata}, which is a function of the five pMSSM model parameters and additional nuisance parameters.
The dimensionality of the likelihood can be reduced to one or two parameters by maximising the likelihood function over the remaining parameters.
The resulting function is called the profile likelihood.

For example, for a single parameter of interest $\theta_i$ and other undesired parameters $\Psi = \{\theta_1,...,\theta_{i-1},\theta_{i+1},...,\theta_{n}\}$ the 1D profile likelihood is defined as:
\begin{equation}
\like(\theta_i)= \max_{\Psi}\like(\theta_i, \Psi) = \like(\theta_i, \hat{\hat{\Psi}}),
\end{equation}
where $\like(\theta_i, \Psi)$ is the likelihood function and $\hat{\hat{\Psi}}$ is the conditional maximum likelihood estimate (MLE) of $\Psi$ for a given $\theta_i$.

Confidence intervals/regions from the resulting 1D/2D profile likelihood maps are determined by adopting the usual Neyman construction with the profile likelihood ratio $\lambda(\theta_i)$ as the test statistic:
\bege
\lambda(\theta_i) = \frac{\mathcal L (\theta_i, \hat{\hat{\Psi}})}{\mathcal L (\hat{\theta_i}, \hat{\Psi})},
\ene
where $\hat{\theta_i}$ and $\hat{\Psi}$ are the unconditional MLEs.

Intervals, or regions, corresponding to 68\%, 95\% and 99\% CL can be estimated by assuming $-2 \ln \lambda(\theta_i)$ is $\chi^2$-distributed which is motivated by Wilks' theorem~\cite{wilks1938}.
This test statistic is used to select the models of interest in this analysis.
For each of the final distributions in Section~\ref{sec:Results} the models included are those within the 95\% confidence interval/region of the profile likelihood.

The software used to sample the parameter space is \SB-v2.0, which is interfaced with the publicly available code \multinest\ v2.18~\cite{Feroz:2007kg,Feroz:2008xx}, an implementation of the nested sampling algorithm~\cite{skilling}.
 This is an updated and improved version of the publicly available \SB\ scanning package~\cite{deAustri:2006jwj,Trotta:2008bp}.
This Bayesian algorithm, originally designed to compute a model's likelihood and to accurately map out the posterior distribution,  can also reliably evaluate the profile likelihood, given appropriate settings~\cite{Feroz:2011bj}.

\SB-v2.0 is interfaced with the following programs:
\textsc{SOFTSUSY}~3.3.10~\cite{Allanach:2001kg} for SUSY spectrum calculations;
\textsc{MicrOMEGAs}~2.4~\cite{Belanger:2006is} to compute the abundance of dark matter;
\textsc{DarkSUSY}~5.0.5~\cite{Gondolo:2004sc} for the computation of $\sigmaSI$, the spin-independent (SI) $\ninoone$-nucleon scattering cross-section, and $\sigma^{\textrm{SD}}_{\chi p}$, the spin-dependent (SD)  $\ninoone$-proton scattering cross-section;
\textsc{SuperIso}~3.0~\cite{Mahmoudi:2008tp} to  compute flavour-physics observables;
and \textsc{SusyBSG}~1.6~\cite{Degrassi:2007kj} for the determination of $\brbsgamma$.
For the computation of the electroweak precision observables described below, the complete one-loop corrections and the available MSSM two-loop corrections have been implemented, as have the full Standard Model results~\cite{Heinemeyer:2007bw}.

Uncertainties in the measured value of the top quark mass, $m_t = 172.99 \pm 0.91\GeV$~\cite{TOPQ-2013-02}, can have a significant impact on the results of SUSY analyses.
Therefore $m_t$ is included as a nuisance parameter in the scans, with a Gaussian prior, in addition to the model parameters described above.
Uncertainties in other Standard Model parameters, as well as astrophysical and nuclear physics quantities that enter the likelihood for the direct-detection experiments (described in Section~\ref{sec:signalLH}), have a very limited impact on the scan.
Thus to limit the dimensionality of the parameter space considered, these other nuisance parameters are fixed in the analysis.
The values used for all Standard Model, astrophysical and hadronic parameters are shown in Table~\ref{tab:nuis_params}.

\begin{table}
\centering
\footnotesize{
\begin{tabular}{l l l |  l l l}
\toprule
\multicolumn{3}{c}{Standard Model} & \multicolumn{3}{|c}{Hadronic} \\
\midrule
$m_t$ [\GeV] & $172.99 \pm 0.91$  &   \cite{TOPQ-2013-02} & $f_{T_u}$ & $0.0457 \pm 0.0065$   & \cite{Ren:12a} \\
$m_b(m_b)^{\overline{\mathrm{MS}}}$ [\GeV] & $4.18\pm 0.03$  & \cite{pdg13} & $f_{T_d}$ & $0.0457\pm 0.0065$  & \cite{Ren:12a}\\
$\left[\alpha_{\mathrm{EM}}(\mZ)^{\overline{MS}}\right]^{-1}$ & $127.944 \pm 0.014$  & \cite{pdg13} & $f_{T_s}$ & $0.043  \pm 0.011$  & \cite{Junnarkar:13a} \\
$\alpha_{\mathrm{S}}(\mZ)^{\overline{MS}}$ & $0.1185 \pm  0.0006$  & \cite{pdg13} & $\Delta_u$ & $0.787 \pm 0.158$   &  \cite{Bali:12a} \\
\cline{1-3}
\multicolumn{3}{c|}{Astrophysical} & $\Delta_d$ & $-0.319 \pm 0.066$  & \cite{Bali:12a} \\
\cline{1-3}
$\rho_{\loc}$ [$\GeV\, {\rm cm}^{-3}$] & $0.4\pm 0.1$ &  \cite{Pato:2010zk} & $\Delta_s$ & $-0.020 \pm 0.011$  & \cite{Bali:12a} \\
$\vs$ [${\rm km \, s}^{-1}$] & $230.0 \pm 30.0$ & \cite{Pato:2010zk} & & &\\
\bottomrule
\end{tabular}
}
\caption{Standard Model, astrophysical and hadronic parameters used in the analysis.
The standard deviation gives the scale of the uncertainty in each (although this is not used in the analysis except in the case of $m_t$).
The astrophysical quantities are the local dark matter density, $\rho_\loc$, and the velocity of the Sun relative to the Galactic rest frame $\vs$.
For the dark matter velocity distribution the so-called Maxwellian distribution is used.
The velocity dispersion is assumed to be $v_{\textnormal{d}} =\sqrt{3/2} \, \vs$.
The hadronic matrix elements, $f_{T_u}$, $f_{T_d}$ and $f_{T_s}$ parameterise the contributions of the light quarks to the proton composition for spin-independent cross-section while $\Delta_{u}$, $\Delta_{d}$ and $\Delta_{s}$ the contributions of the light quarks to the total proton spin for the spin-dependent neutralino--proton scattering cross-section.
\label{tab:nuis_params}}
\end{table}

\subsection{Experimental constraints in the initial likelihood scan}
\label{sec:nonsusydata}
A set of existing experimental constraints is used in the initial likelihood scan over the 5D pMSSM to select the models in which to consider the impact of the ATLAS SUSY searches. They are implemented with a joint likelihood function,  whose logarithm takes the following form:
\begin{equation}
\ln \likeJ =  \ln \like_\text{EW}+ \ln \like_\text{B} + \ln \like_{\Ohsq}  + \ln \like_\text{DD} + \ln \like_\text{Higgs} + \ln \like_{\text{LEP-}\chinoonepm},
\end{equation}
where $\like_\text{EW}$ represents electroweak precision observables, $\like_\text{B}$ $B$-physics constraints, $\like_{\Ohsq}$ measurements
of the cosmological dark matter relic density, $\like_\text{DD}$ direct dark matter detection constraints, $\like_\text{Higgs}$ the ATLAS measurement of the
Higgs boson mass, and $\like_{\text{LEP-}\chinoonepm}$ the LEP2 limit on the chargino mass.

Table~\ref{tab:exp_constraints} shows the set of experimental constraints used in the analysis.
Their implementation is summarised below.

\begin{table}
\centering
\footnotesize{
\begin{tabular}{l | S l l | l}
\toprule
Observable & \multicolumn{1}{c}{Mean value} & \multicolumn{2}{c|}{Standard deviation} & Ref. \\
 & & Experimental & Theoretical & \\\midrule
\mW [$\GeV$] & 80.385 & 0.015 & 0.01 & \cite{Agashe:2014kda} \\
\swel & 0.23153 & 0.00016 & 0.00010 & \cite{ALEPH:2005ab} \\
\GZ [$\GeV$] & 2.4952 & 0.0023 & 0.001 & \cite{lepwwg} \\
$\Gamma^{\mathrm{inv}}_{Z}$ [$GeV$] & 0.499 & 0.0015 & 0.001 & \cite{Agashe:2014kda} \\
$\sigma^0_{\mathrm{had}}$ [nb] & 41.540 & 0.037 & - & \cite{lepwwg} \\
$R^0_{\ell}$ & 20.767 & 0.025 & - & \cite{lepwwg} \\
$R^0_b$ & 0.21629 & 0.00066 & - & \cite{lepwwg} \\
$R^0_c$ & 0.1721 & 0.003 & - & \cite{lepwwg} \\
$\brbsgamma \times 10^4$ & 3.55 & 0.26 & 0.30 & \cite{Agashe:2014kda}\\
$\RBtaunu$   &  1.62  & 0.57  & - & \cite{Akeroyd:2010qy}  \\
$\brbsmumu \times 10^{9} $ & 2.9 & 1.1 & 0.38 & \cite{bsmm}\\
$\Omega_\chi h^2$ & 0.1186 & 0.0031 & 0.012 & \cite{Ade:2013zuv} \\
\mh [$\GeV$] & 125.36  & 0.41  & 2.0 & \cite{HIGG-2013-12} \\
\midrule
   &  \multicolumn{1}{c}{Limit}   & \multicolumn{2}{r|}{} & Ref. \\ \midrule
$m_\chi$ vs.\ $\sigmaSI$ & \multicolumn{3}{l|}{XENON100 2012  ($224.6 \times 34\,{\rm kg\,days}$) } & \cite{Aprile:2012nq} \\
$m_\chi$ vs.\ $\sigma^{\textrm{SD}}_{\chi p}$  & \multicolumn{3}{l|}{XENON100 2012  ($224.6 \times 34 \,{\rm kg\,days}$) } & \cite{Aprile:2013doa} \\
$m_\chi$ vs.\ $\sigmaSI$  & \multicolumn{3}{l|}{LUX 2013  ($118 \times 85.3\,{\rm kg\,days}$) } & \cite{Akerib:2013tjd} \\
Chargino mass  &  \multicolumn{3}{c|}{LEP2}  & \cite{Agashe:2014kda}\\
\bottomrule
\end{tabular}
}
\caption{Summary of experimental constraints that are used in the likelihood.
Upper part: measured observables, modelled with a Gaussian likelihood with the standard deviation $(\sigma^2+\tau^2)^{1/2}$, where $\sigma$ is the experimental and $\tau$ the theoretical uncertainty.
Lower part: observables for which only limits currently exist.
$\sigmaSI$ and $\sigma^{\textrm{SD}}_{\chi p}$ denote spin-independent and spin-dependent LSP--nucleon scattering cross-sections respectively.
See text for further information about the explicit form of the likelihood function. All the observables are described in Section 3.
\label{tab:exp_constraints}}
\end{table}

The constraints on the electroweak precision observables are obtained from \Zboson-pole measurements at LEP~\cite{lepwwg}, and include the constraint on the effective electroweak mixing
angle for leptons \swel, the total width of the \Zboson\ boson \GZ, the invisible \Zboson\ boson width   $\Gamma^{\mathrm{inv}}_{Z}$,
the hadronic pole cross-section $\sigma^0_{\mathrm{had}}$, as well as the decay width ratios $R^0_l$, $R^0_b$ and $R^0_c$.
The combined Tevatron and LEP \Wboson\ boson mass (\mW) estimate ~\cite{Agashe:2014kda} is also included.
The $B$-physics constraints include a number of world averages obtained by the Heavy Flavour Averaging Group, including the branching fraction $\brbsgamma$ and the ratio of the branching fraction of the decay $\Bu \to \tau \nu$ to its branching fraction predicted in the Standard Model~\cite{Agashe:2014kda}.
Finally, the measurement of the rare decay branching fraction $\brbsmumu$ from the LHCb experiment at the LHC is used~\cite{bsmm}.
The electroweak precision and $B$-physics constraints are applied as Gaussian likelihoods with means and standard deviations as indicated in Table~\ref{tab:exp_constraints}.

For the cosmological constraints the Planck Collaboration's constraint on the dark matter relic abundance is used, but it is implemented differently depending on the proportion of dark matter attributed to neutralinos.
If the neutralino were to make up all of the dark matter in the universe, the result from Planck temperature and lensing data, $\Ohsq = 0.1186 \pm 0.0031$, would be applied as a Gaussian likelihood~\cite{Ade:2013zuv}.
But here, the neutralino is allowed to be a sub-dominant dark matter component, and the Planck relic density measurement is instead applied as an upper limit.
The effective likelihood for the upper limit, taking into account the error, is given by the expression
\bege \label{eq:upperbound}
\like_{\Ohsq} = \like_0 \int_{\OhLSP/\siW}^\infty
\mathrm{e}^{-\frac{1}{2}(x-r_\star)^2} x^{-1}\mathrm{d}x ,
\ene
as derived in the appendix of Ref.~\cite{Bertone:2010ww}. $ \like_0$ is an irrelevant normalisation constant,  $r_\star \equiv \muW/\siW$, and $\OhLSP$ is the predicted relic density of neutralinos as a function of the model parameters. Here $\muW$ refers to the value of $\OhLSP$ inferred by the Planck Collaboration and $\siW$ to its uncertainty. Both numbers are given in Table 4.
A fixed theoretical uncertainty, $\tau = 0.012$, is also added in quadrature to the experimental error, in order to account for the numerical uncertainties entering in the calculation of the relic density from the SUSY parameters.

When neutralinos are not the only constituent of dark matter, the rate of events in a direct-detection experiment is proportionally smaller, as the local neutralino density,  $\rho_\chi$,  is now smaller than the total local dark matter density, $\rho_\text{DM}$.
The suppression is given by the factor $\xi \equiv \rho_\chi / \rho_{\mathrm{DM}}$.
Following Ref.~\cite{Bertone:2010rv}, the ratio of local neutralino density to total dark matter densities is assumed to be equal to that for the cosmic abundances, thus a scaling ansatz is adopted:
\bege \label{eq:scaling_ansatz}
\xi \equiv \frac{\rho_\chi}{\rho_{\mathrm{DM}}} = \frac{\Omega_\chi}{\Omega_{\mathrm{DM}}}.
\ene
For $\Omega_{\mathrm{DM}}$, the central value measured by the Planck Collaboration, $\Omega_{\mathrm{DM}}h^2 = 0.1186$, is used~\cite{Ade:2013zuv}.

The direct-detection constraint uses the recent results from XENON100, with 225 live days of data collected between February 2011 and March 2012 with a $34\,{\rm kg}$ fiducial volume~\cite{Aprile:2012nq}.
The treatment of XENON100 data is described in detail in Ref.~\cite{Bertone:2011nj}.
The likelihood function is built as a Poisson distribution for observing $N$ recoil events when $N_{\mathrm{s}}(\params)$ signal plus $N_{\mathrm{b}}$ background events are expected.
The expected number of background events the XENON100 run is $N_{\mathrm{b}} = 1.0 \pm 0.2$, while the collaboration reported $N=2$ events observed in the pre-defined signal region.
An updated version of the likelihood function described in Refs.~\cite{Bertone:2011nj,Strege:2011pk} is used.
In addition the LUX data~\cite{Akerib:2013tjd} was included using the likelihood computed by the LUXCalc
package~\cite{Savage:2015xta}.
The likelihood is constructed from a Poisson distribution in which the numbers of observed and background events are 1 and 0.64, respectively.

For the implementation of the Higgs boson likelihood the most recent measurement by the ATLAS experiment of the mass of the Higgs boson is used, $\mh = 125.36 \pm 0.37 \pm 0.18\GeV$, where the first error is statistical and the second error is systematic~\cite{HIGG-2013-12}.
A theoretical error of $2\GeV$~\cite{Allanach:2004rh} is added in quadrature to these uncertainties.
The observed upper limit of 0.23 on the branching fraction for Higgs boson decays into invisible particles~\cite{HIGG-2015-03} (e.g. $\nu, \ninoone$) is not included. Including this bound would exclude at the 95\% confidence level (CL) 5\% of models surviving the initial likelihood scan, and 8\% of those remaining after the electroweak SUSY analysis constraints have been applied.

Finally, the likelihood associated with the $m(\chinoonepm)$ constraint from LEP2 data is taken
from Equation~(3.5) of Ref.~\cite{deAustri:2006pe}, where an experimental lower bound of $92.4\GeV$~\cite{Agashe:2014kda}
and a theoretical uncertainty of 5\% from the \textsc{SOFTSUSY}~3.3.10 prediction of the spectrum is assumed.

\subsection{Phenomenology of the LSP}

As mentioned in Section~\ref{sec:scanning}, the results of the likelihood scan are used to select models upon which to consider the sensitivity of the electroweak SUSY searches. Figure~\ref{fig:resLSPcomposition} displays the LSP composition of those models within the 95\% CL 2D contours, and their distribution in the $\ninoone$ versus $\chinoonepm$ mass plane.
The colours encode the $\ninoone$ composition of the models.
Three distinct regions are seen, which correspond to different mechanisms to enhance the annihilation cross-section and thus avoid having a cosmological relic density larger than observed.
There is the so-called \Zboson-funnel region, where the LSP mass is close to $45\GeV$ and it is  mostly bino-like.
In this case, the annihilation rate is proportional to the higgsino fraction of the $\ninoone$.
The region centred on $m(\ninoone) \sim 60\GeV$ corresponds to a $\ninoone$ that annihilates through a mechanism similar to that in the \Zboson-funnel but involving the lightest Higgs boson instead.
This is the so-called $h$-funnel, and the annihilation rate is proportional to the higgsino fraction as well as the combined bino and wino fraction.
In each funnel, the $\ninoone$ annihilation rate is enhanced due to a pole in the propagator ($2 m(\ninoone) \sim m_Z$ or $m_h$, respectively) and thus the Planck constraint can be satisfied.
Finally, there is a compressed region, where $m(\ninoone)\approx m(\chinoonepm)$.
Here, the LSP composition is less constrained --- in particular, higgsino-like and wino-like states are likely, as well as wino--higgsino mixed states.
Some model points with $m(\ninoone) \gtrsim 200\GeV$ have a non-compressed spectrum and a nearly pure bino-like LSP.
These correspond to the so-called $A$-funnel region, where dark matter annihilates through the pseudoscalar Higgs boson pole.

\begin{figure}[htbp!]
\centering
\includegraphics[width=0.6\linewidth]{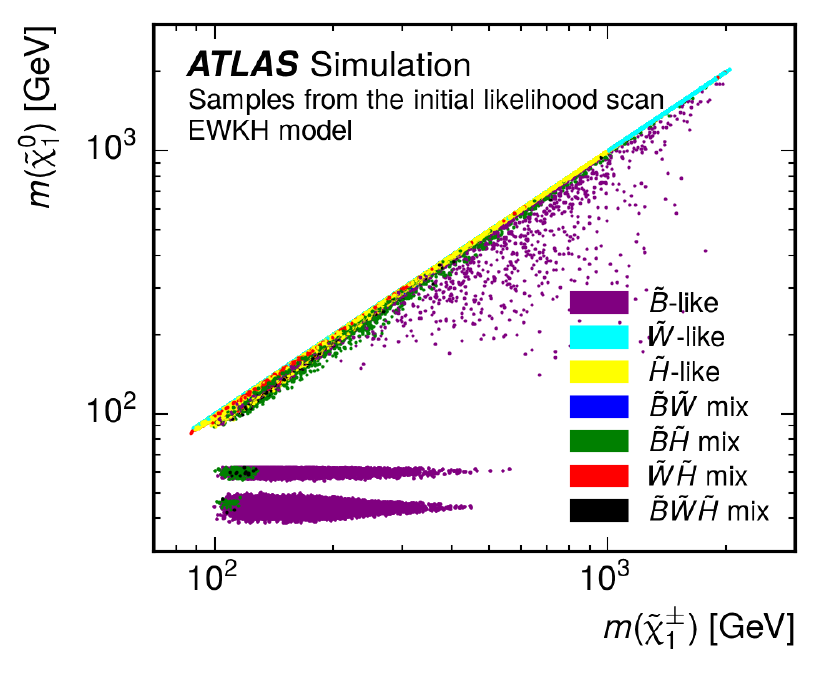}
\caption{
Scatter plot of models in the $m(\ninoone)$ vs.\ $m(\chinoonepm)$ plane with the colour encoding which category of $\ninoone$ composition the model belongs to.
The $\ninoone$ is defined as bino-like ($\tilde{B}$-like), wino-like ($\tilde{W}$-like) or higgsino-like ($\tilde{H}$-like) if the relevant fraction is at least $80\%$.
A mixed $\ninoone$ has at least $20\%$ of each denoted component and $< 20\%$ of any other component.
The models considered are all within the 95\% confidence region found using the initial likelihood scan.
}
\label{fig:resLSPcomposition}
\end{figure}

\section{Signal simulation and evaluation of ATLAS constraints}
\label{sec:signalLH}

Constraints from ATLAS SUSY searches are imposed on the $570\,599$ models generated in the initial likelihood scan by generating and simulating events from a subset of these models.
The models are split into three categories: those considered to be already excluded by pre-existing constraints and having a $\ninoone$ lighter than $1\TeV$ ($108\,740$ models); those where the considered analyses are assumed to be insensitive without performing a detailed analysis ($134\,624$ models); and those that are simulated to assess the impact of the searches in Table~\ref{tab:susySearches} ($326\,951$ models).

The pre-existing constraint defining the first category of models is the LEP2 limit on the mass of the lightest chargino, $m(\chinoonepm)>92.4\GeV$.
The second category, consisting of models for which the considered searches are not expected to have any sensitivity, is defined by estimating the total production cross-section for SUSY particle production, using {\sc Prospino}2~\cite{Beenakker:1996chA,Beenakker:1997utA,Beenakker:1999xhA,Spira:2002rd,Plehn:2004rp}.
The searches are not optimised for detecting the decay products of sparticles very close in mass to the LSP, and therefore a process $pp\to\tilde{\chi}_i\tilde{\chi}_j$ is only included in the cross-section calculation if $\Delta m(\tilde{\chi}_i,{\mathrm{LSP}})$ or $\Delta m(\tilde{\chi}_j,{\mathrm{LSP}})$ is greater than $5\GeV$.
Models with a total cross-section for all considered electroweak SUSY production processes below $0.25\,{\rm fb}$ are placed in the second category and not processed further at this stage.
They are, however, included as unexcluded models in Section~\ref{sec:Results}.

The remaining $326\,951$ models, in the third category, are simulated at particle level using \MADGRAPH~1.5.12 \cite{Alwall:2007st} with the CTEQ 6L1 parton density function set~\cite{Pumplin:2002vw} and \PYTHIA~6.427~\cite{Sjostrand:2006za} with the AUET2B~\cite{ATL-PHYS-PUB-2011-009} set of tuned parameters.
\MADGRAPH\ is used to generate the initial pair of sparticles and up to one additional parton, while \PYTHIA\ is used for all sparticle decays and parton showering.
\TAUOLA~\cite{Jadach:1993hs} and \PHOTOS~\cite{Golonka:2005pn} are used to handle the decays of $\tau$-leptons and the final-state radiation of photons, respectively.
Expected signal region yields are calculated for each of the four considered analyses using these simulated events.

To avoid the computational cost of processing every model with the ATLAS detector simulation, a ``calibration procedure'' is used to extract \cls\ values for the models using the particle-level signal region yields described above.
Of the $326\,951$ simulated models, a random sample of 500 models was selected and processed using a fast \texttt{GEANT4}-based \cite{Agostinelli:2002hh} simulation of the ATLAS detector, with a parameterisation of the performance of the ATLAS electromagnetic and hadronic calorimeters \cite{SOFT-2010-01} and full event reconstruction. The selected models follow approximately the initial likelihood scan and thus span the relevant parameter space.
The number of events generated for each of these models corresponds to approximately four times the recorded integrated luminosity collected at $\sqrt{s} = 8\,\TeV$, i.e.\ $80\,\ifb$.
For these simulated models, signal cross-sections are calculated at next-to-leading (NLO) order in the strong coupling constant using {\sc Prospino}2~\cite{Beenakker:1999xhA}.
These cross-sections are in agreement with the NLO calculations matched to resummation at the next-to-leading-logarithmic accuracy (NLO+NLL) within $\sim2\%$~\cite{Fuks:2012qx,Fuks:2013vua,Fuks:2013lya}.
The nominal cross-section and the uncertainty are taken from an envelope of cross-section predictions using different parton distribution function sets and factorisation and renormalisation scales, as described in Ref. ~\cite{Kramer:2012bx}.

These 500 models are then analysed using the full statistical framework~\cite{Baak:2014wma} of the original ATLAS electroweak SUSY analyses and a \cls\ value is calculated for each of them.
One difference with respect to the published analyses is that signal regions that would normally be statistically combined in the likelihood fit are now treated as separate signal regions, and \cls\ values are calculated for each region.
Similarly, for binned signal regions each bin is treated separately.
The results from the 500 models are used to fit a ``calibration function'' between the particle-level yields and the \cls\ values for each signal region.
This accounts for the SM background prediction in each signal region, together with the observed data. There is one remaining free parameter, which roughly corresponds to the average selection efficiency for SUSY events that pass the particle-level selection. Only those signal regions where the average efficiency could be determined with a statistical precision of better than 20\% are considered in the final analysis.
In addition, it is required that at least one of the 500 models is excluded, with expected and observed $\cls < 0.05$.
Of the original 44 signal regions, 25 pass these requirements.
The 19 rejected signal regions typically have a low acceptance for the EWKH models, due to either very stringent kinematic criteria, or a requirement for \tauhad\ candidates, which have a low yield in the EWKH models considered, due to the very high mass of the stau.
The real selection efficiency varies from model to model, and the calibration procedure therefore can only gives accurate results when averaged over many models. No additional systematic uncertainty for model-to-model variations is applied.

This simplified method provides an efficient way to calculate the impact of the electroweak searches and the calibration functions are used to extract \cls\ values for all $326\,951$ considered models.
The best constraints on any signal model would be obtained from a statistical combination of all relevant signal regions; however, this is not possible with this simplified approach so instead a conservative approach is used where the \cls\ value is taken from the signal region with the smallest expected \cls\ value.

\section{Impact of the ATLAS electroweak SUSY searches}
\label{sec:Results}
In this section the impact of the ATLAS electroweak SUSY searches is discussed in terms of 1D and 2D distributions.
The models considered for each distribution are those within the 95\% confidence region according to the initial likelihood scan outlined in Section~\ref{sec:theoreticalFramework}.
There are $438\,589$ and $472\,933$ such models in the 1D and 2D case, respectively.

A model is considered to be excluded by the ATLAS electroweak SUSY searches if the observed \cls\ value, calculated as explained in Section~\ref{sec:signalLH}, is less than 0.05.
For the 1D distributions in this section, stacked plots are used to indicate the contributions of the $2\ell$, $3\ell$ and $4\ell$ searches.
The $2\tau$ search is found to be insensitive, relative to the other searches, due to the lack of light staus in these models.
Signal regions of the $3\ell$ and $4\ell$ searches that require \tauhad\ candidates are similarly insensitive to these models.
If more than one search can exclude a model, the one with the smallest expected \cls\ value is chosen, following the procedure in Section~\ref{sec:signalLH}.
For the 2D plots the colours represents the fraction of models which are excluded by ATLAS data at $95\%$ CL. In all of the distributions the fractions displayed correspond to the proportion of models excluded for a given bin in the parameter space.

Of the $472\,933$ models within the two-dimensional 95\% CL bound before the ATLAS electroweak SUSY analyses are considered, approximately 3\% are excluded by the searches considered (listed in Table~\ref{tab:susySearches}).
The $3\ell$ search is the most powerful of the four analyses across these models, having the signal region with the lowest expected \cls\ for $63.3\%$ of the excluded models.
The high sensitivity of this search is largely due to a signal region that is binned in kinematic quantities such as the dilepton invariant mass and \met (the signal region is called SR0$\tau$a in Ref.~\cite{SUSY-2013-12}).
The 20 bins of SR0$\tau$a are treated here as 20 individual signal regions, the most powerful of which (for these models) is bin 16, requiring a \Zboson boson candidate and stringent lower limits on the transverse mass (\mt) and \met.
The $2\ell$ and $4\ell$ searches exclude smaller fractions of models, although they have areas of unique sensitivity, as discussed below.

\subsection{Impact on the electroweakino masses}
\label{sec:res1Dmasses}

The fractions of models excluded as a function of $m(\ninoone)$, $m(\chinoonepm)$, and $m(\ninotwo)$ are shown as 2D and 1D distributions in Figures~\ref{fig:res2D} and~\ref{fig:res1Dmasses}, respectively.
Areas where no models survive the initial likelihood scan are left white in Figure~\ref{fig:res2D} and Figure~\ref{fig:res1Dmasses}.
For example, chargino masses below $100\GeV$ are strongly disfavoured due to the LEP2 constraint, which also impacts the range of \ninotwo masses that can be considered.
The \Zboson- and $h$-funnel regions are also clearly visible in both Figures~\ref{fig:resC1vN1} and~\ref{fig:resN1}.

These results show that the considered searches effectively constrain the \Zboson- and $h$-funnel regions of the parameter space, with the greatest impact when $m(\chinoonepm) \lesssim 300\GeV$.
In this scenario the leptons produced in the decay of the produced electroweakinos to the LSP have a large signal acceptance, and the production cross-section of wino- and higgsino-like particles can reach $\mathcal{O}({\rm pb})$ with these masses.
The searches have a negligible impact in the compressed region where $m(\ninoone)\approx m(\chinoonepm)$, since the reconstruction efficiency of low-$\pt$ leptons ($\pt \lesssim 5 \GeV$) is small.

Overall, the results are dominated by the $3\ell$ search, as explained above.
The $4\ell$ search is uniquely sensitive to a small fraction of models in a particular region of the parameter space where all of the electroweakinos have masses smaller than approximately $300\GeV$.
These models also have a particular pattern of wino/higgsino mixing that especially favours the SR0Z signal region, which requires a \Zboson candidate and significant \met~\cite{SUSY-2013-13}.
The signal process $pp\to\ninotwo\ninothree\to\Zboson\ninoone\Zboson\ninoone$ was already considered in the $4\ell$ search paper as a simplified model; however, the relatively light ($m \lesssim 300\GeV$) wino-like \ninofour and \chinotwopm particles supplement the search sensitivity via long cascades such as $\chinotwop\chinotwom\to(\Zboson\chinoonep)(\Wminus\ninotwo)\to(\Zboson\Wplus\ninoone)(\Wminus\Zboson\ninoone)$.
The $2\ell$ search is mainly used to exclude models with extremely light higgsino-like particles ($m(\chinoonepm,\ninotwo)\sim 100$--$130\GeV$), with a bino-like LSP in the \Zboson- or $h$-funnel region.
The exclusion power arises mostly from the signal region SR-WWa, which is optimised for processes such as $pp\to\chinoonep\chinoonem\to(\Wplus{}^*\ninoone)(\Wminus{}^*\ninoone)$ where $m(\chinoonepm)-m(\ninoone) < m_W$~\cite{SUSY-2013-11}.
The wino-like electroweakinos are usually significantly more massive ($m(\ninofour,\chinotwopm)\gtrsim 300\GeV$), such that the search is mainly sensitive to \chinoonep{}\chinoonem, \ninotwo{}\chinoonepm and \ninothree{}\chinoonepm pair production.

Comparing Figures~\ref{fig:resC1vN2} and~\ref{fig:resN2} shows that these searches are in general only sensitive to models where the \ninotwo mass is smaller than about $300\GeV$.
The proportion of excluded models approaches 30\% in the best case, for $m(\ninotwo) \approx 120\GeV$.
This subset of models corresponds most closely to the canonical signature targeted by the $2\ell$, $3\ell$ and $4\ell$ searches, where wino- or higgsino-like particles decay to a bino-like LSP and either a \Wboson or \Zboson boson (which may be off-shell).
These searches are expected to be less sensitive in the case where the \ninotwo and \chinoonepm masses are not degenerate, as seen in Figure~\ref{fig:resC1vN2}.
Then, even if the \chinoonepm is accessible, typically this implies that it and the LSP are both mostly wino-like, with a very small mass difference that prevents detection by the considered analyses.
The ATLAS search for disappearing tracks~\cite{SUSY-2013-01} targets this kind of signature, in the case where the mass difference is small enough that the \chinoonepm can traverse a significant portion of the detector before it decays ($\Delta m\lesssim 200\MeV$).
A full consideration of this search would lead to further constraints in this part of the parameter space as shown in Ref.~\cite{SUSY-2014-08}.

\begin{figure}[htbp!]
\centering
\subfloat[]{
\includegraphics[width=0.45\linewidth,page=1]{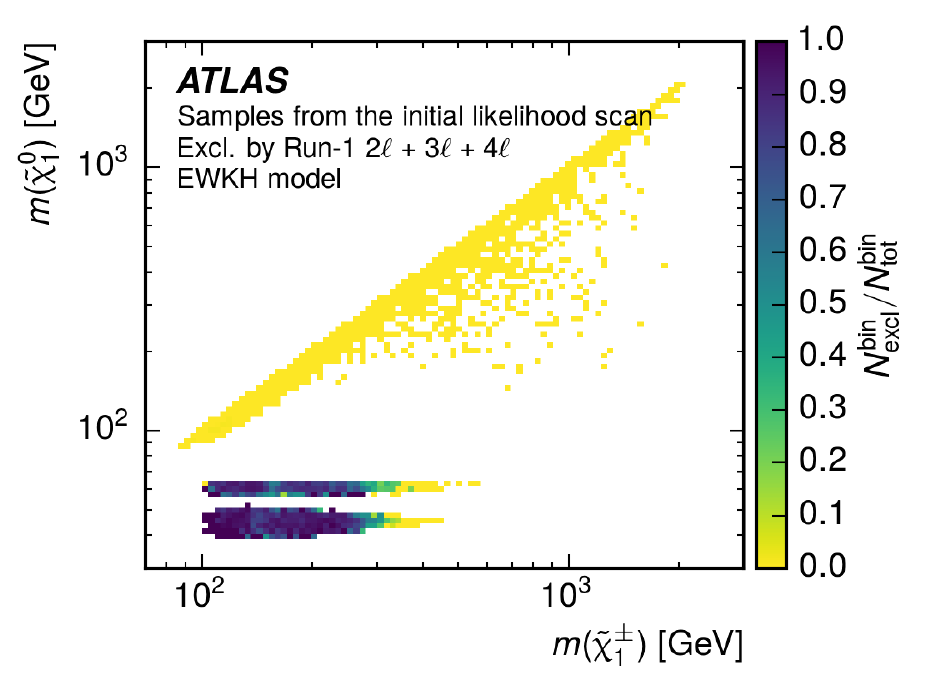}
\label{fig:resC1vN1}
}
\subfloat[]{
\includegraphics[width=0.45\linewidth,page=1]{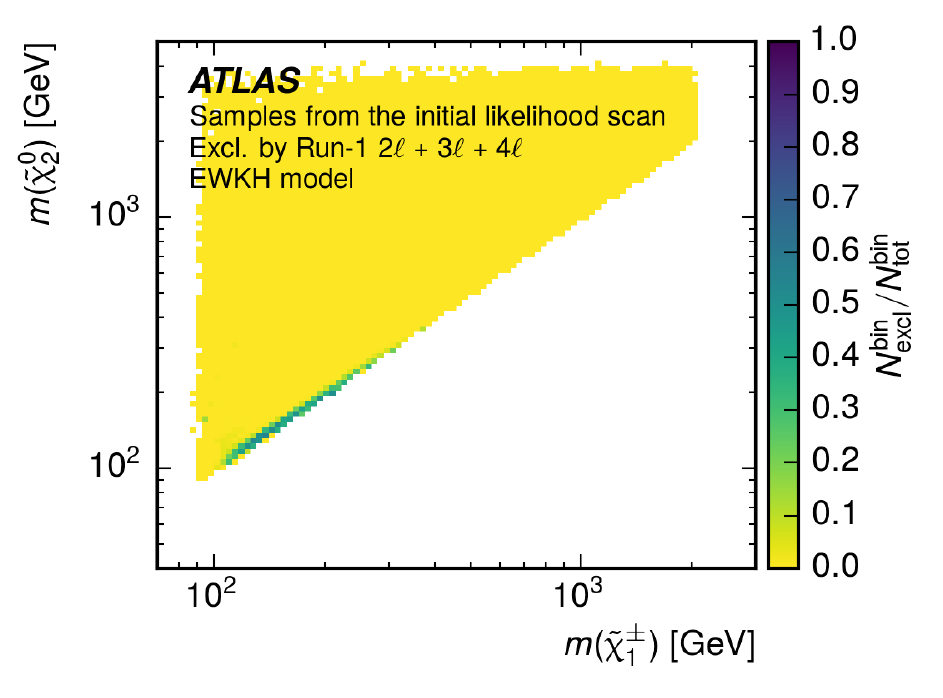}
\label{fig:resC1vN2}
}\\[-2ex]
\caption{
The bin-by-bin fraction of models excluded as a 2D function of sparticle masses.
The colour encodes the fraction of models excluded.
The models considered are all within the 2D 95\% confidence region found using the initial likelihood scan. No such models are in the white regions, and therefore the coloured bins indicate the 95\% CL contours for the initial likelihood scan.}
\label{fig:res2D}
\end{figure}

\begin{figure}[htbp!]
\centering
\subfloat[]{
\includegraphics[width=0.4\linewidth,page=1]{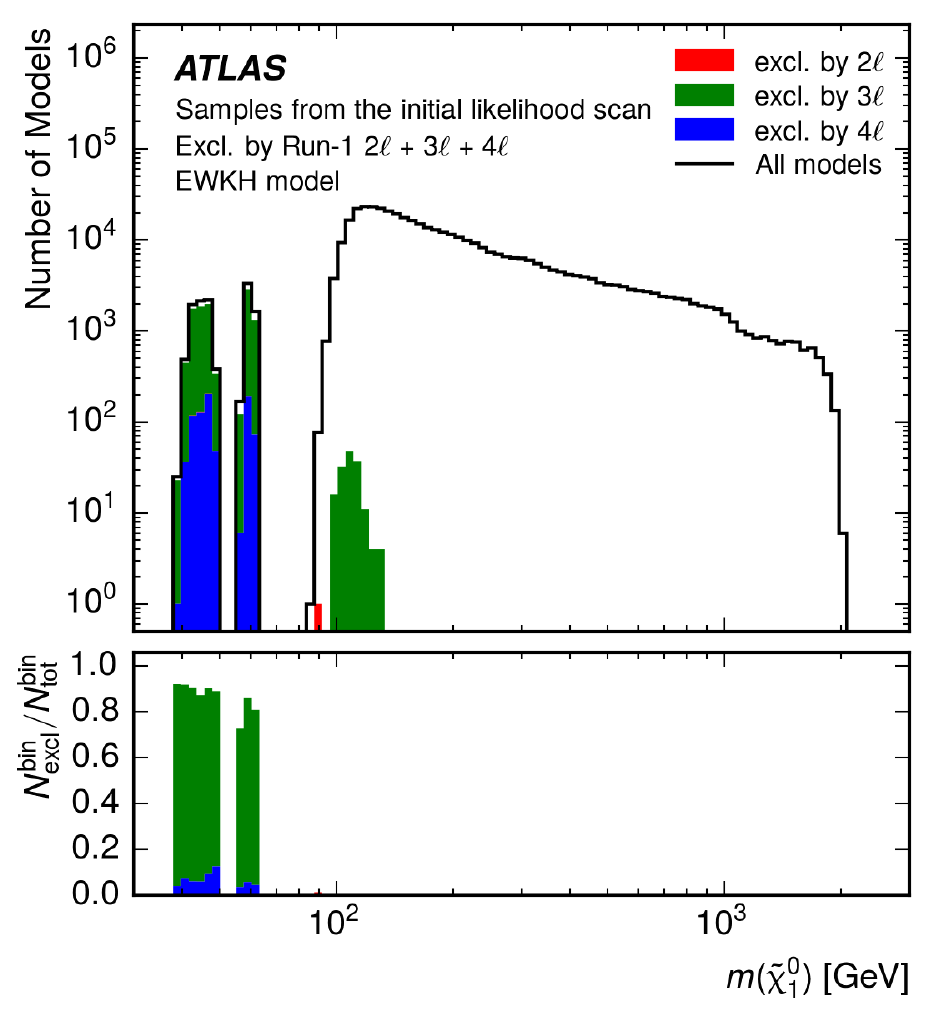}
\label{fig:resN1}
}
\subfloat[]{
\includegraphics[width=0.4\linewidth,page=1]{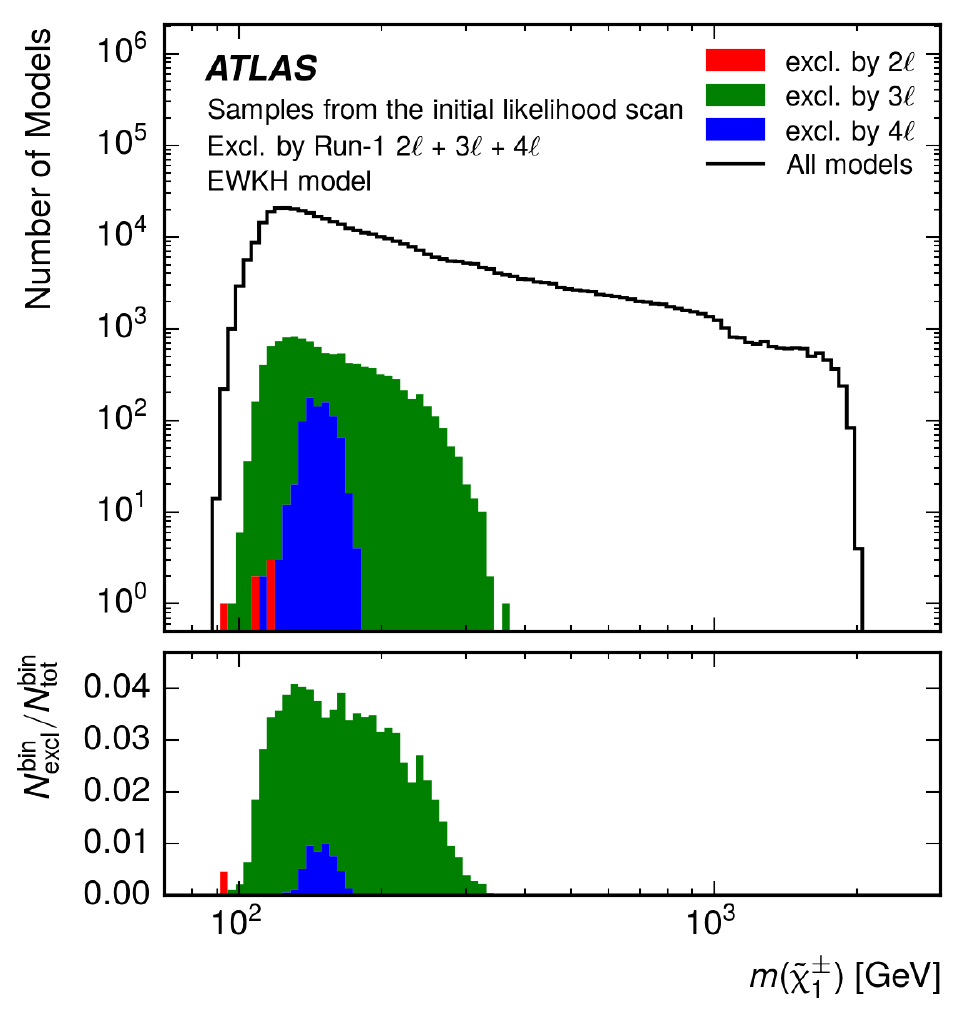}
\label{fig:resC1}
}\\
\subfloat[]{
\includegraphics[width=0.4\linewidth,page=1]{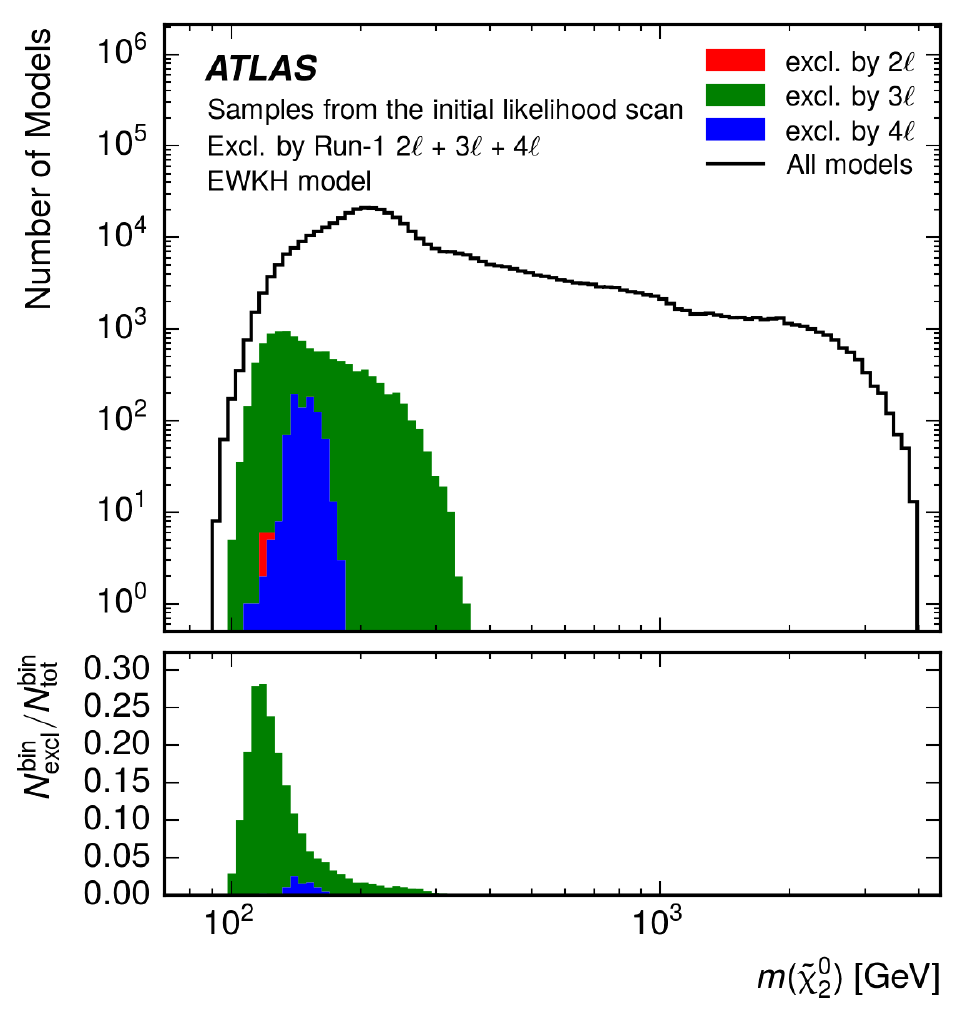}
\label{fig:resN2}
}\\[-1ex]
\caption{
The number of models sampled by the initial likelihood scan, and the stacked bin-by-bin number of models excluded by the Run 1 ATLAS SUSY searches as a 1D function of $m(\ninoone)$, $m(\chinoonepm)$, and $m(\ninotwo)$.
The lower part of each figure shows the fraction of models excluded by the Run 1 ATLAS SUSY searches.
The red bins indicates the fraction that is excluded by a $2\ell$ SR, the green by a $3\ell$ SR, and blue by a $4\ell$ SR.
The models considered are all within the 1D 95\% confidence interval found using the initial likelihood scan.
}
\label{fig:res1Dmasses}
\end{figure}

\subsection{Impact on the EWKH model parameters}
\label{sec:res1Dmodelparam}

Figures~\ref{fig:res2DmodelParam} and~\ref{fig:res1Dmodelparam} display the fraction of models excluded for the five EWKH parameters: $M_1$, $M_2$, $\mu$, $m_A$ and $\tan\beta$.
As before, regions of the parameter space are visible where no models are allowed.
For example, there are no models with $M_2$ or $|\mu|$ less than $80\GeV$ due to the LEP2 constraint on the  $\chinoonepm$ mass.
The measured value of \brbsmumu\ is compatible with the Standard Model prediction, disfavouring the region with $m_A \lesssim 500\GeV$ in Figure~\ref{fig:resMA} as contributions to that process typically scale as $\sim \tan^6 \beta/m_A^2$.
Finally, values of $\tan\beta \gtrsim 10$ (Figure~\ref{fig:resTANB}) are strongly favoured because the tree-level contribution to the Higgs boson mass is maximised.

As seen in Figures~\ref{fig:res2DmodelParam} and~\ref{fig:resM1}--\ref{fig:resMU}, the considered searches have the strongest impact when $|M_1|$, $M_2$ and $|\mu|$ are all small ($\ll 1\TeV$), where the SUSY particle production cross-section is large.
The searches have the strongest impact where the \ninoone is light and bino-like; approximately 86\% of models with $|M_1| < 85\GeV$ are excluded, which corresponds to the region $m(\ninoone)<65 \GeV$ in Figure~\ref{fig:res1Dmasses}. The impact on $M_2$ and $\mu$ is less severe, where the excluded fraction peaks at about 4\%.
In the case of $M_2$, a small number of models with $M_2 > 1\TeV$ are excluded, corresponding to models with a light higgsino spectrum and a bino-like LSP.

The considered searches can only provide indirect constraints on the remaining model parameters, $m_A$ and $\tan\beta$.
Therefore, the features in Figures~\ref{fig:resMA} and~\ref{fig:resTANB} are driven by the properties of models with a low-mass LSP in the \Zboson- or $h$-funnel.
Although the pseudoscalar boson does not enter directly into the phenomenology of the considered electroweak searches, the proportion of excluded models is greatest for values of $m_A$ below $1\TeV$, while the excluded models span a wide range of $\tan\beta$ between about 20 and 50.

\begin{figure}[htbp!]
\centering
\subfloat[]{
\includegraphics[width=0.45\linewidth,page=1]{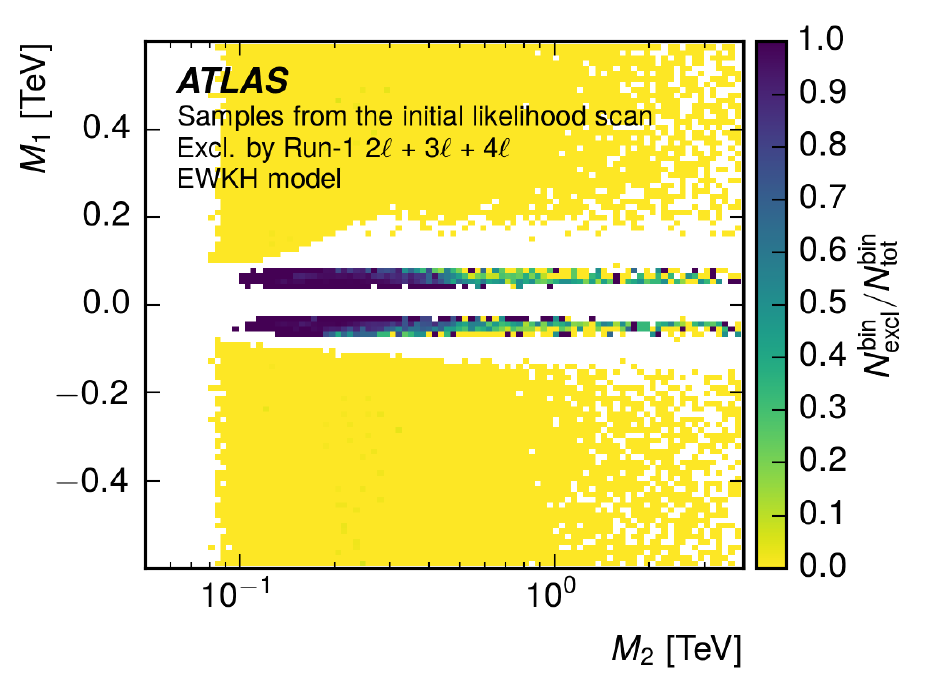}
\label{fig:resM1vM2}
}%
\subfloat[]{
\includegraphics[width=0.45\linewidth,page=1]{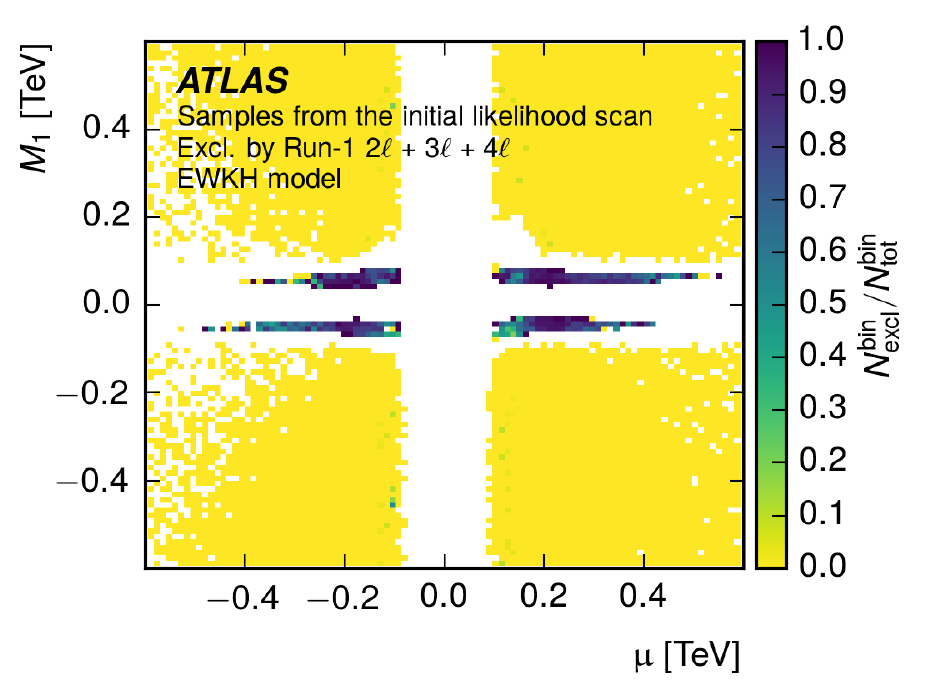}
\label{fig:resM1vMU}
}\\[-1ex]
\caption{
The bin-by-bin fraction of models excluded as a 2D function of model parameters.
The colour encodes the fraction of models excluded.
The models considered are all within the 2D 95\% confidence region found using the initial likelihood scan. No such models are in the white regions, and therefore the coloured bins indicate the 95\% CL contours for the initial likelihood scan.
The plots are truncated in $|M_1|$ and $|\mu|$ to highlight the region of ATLAS electroweak SUSY sensitivity.
}
\label{fig:res2DmodelParam}
\end{figure}

\begin{figure}[htbp!]
\centering
\subfloat[]{
\includegraphics[width=0.35\linewidth,page=1]{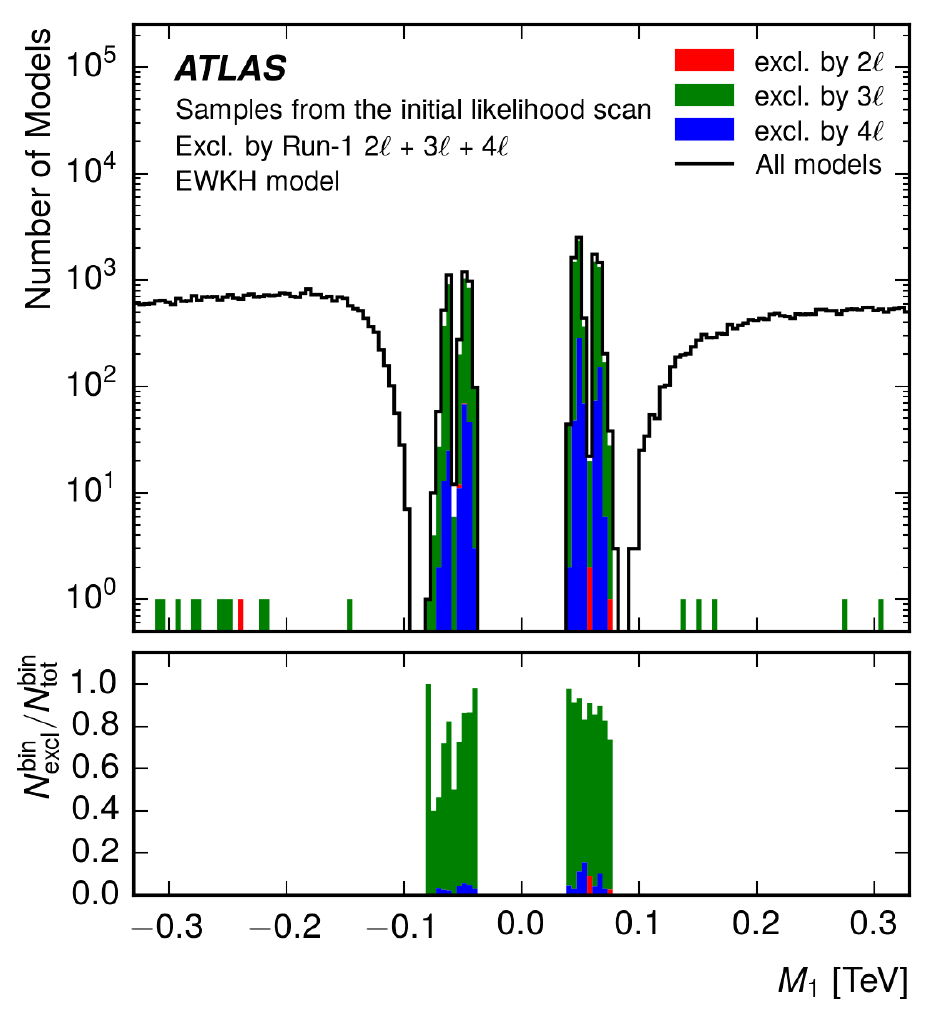}
\label{fig:resM1}
}
\subfloat[]{
\includegraphics[width=0.35\linewidth,page=1]{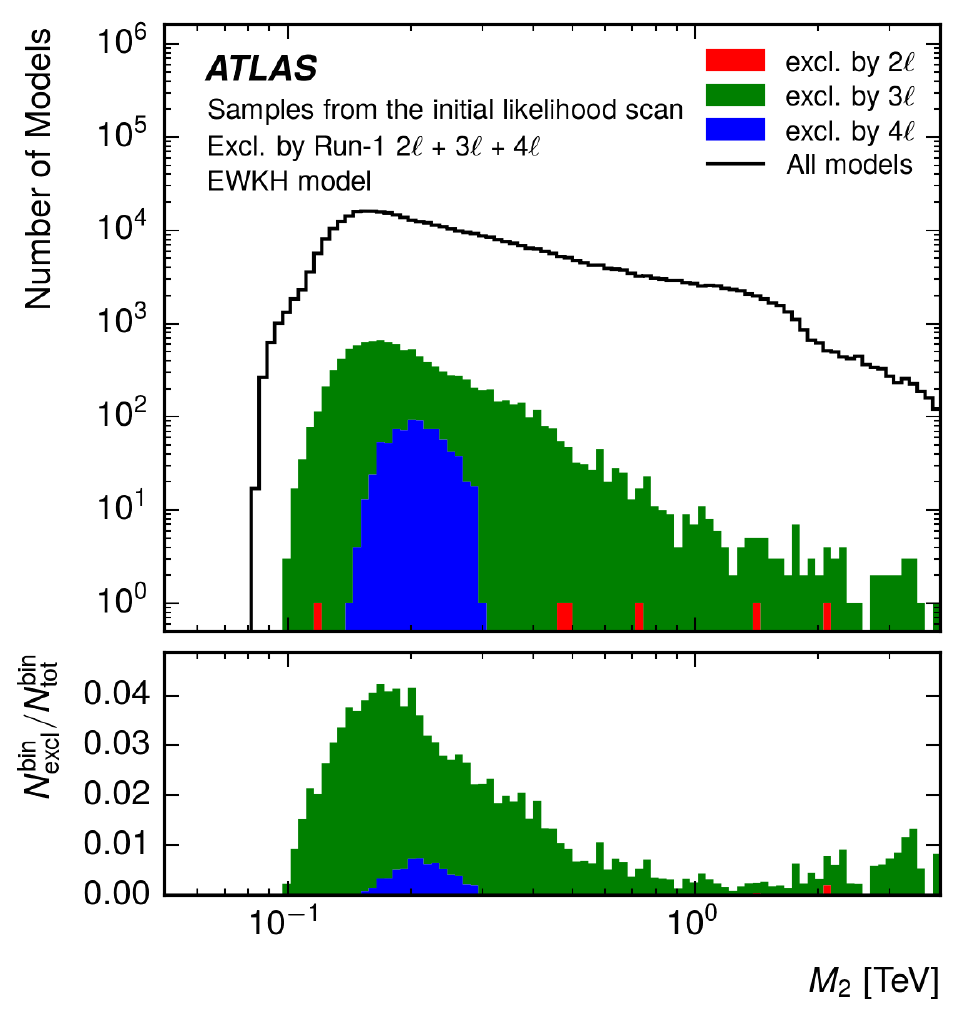}
\label{fig:resM2}
}\\[-2ex]
\subfloat[]{
\includegraphics[width=0.35\linewidth,page=1]{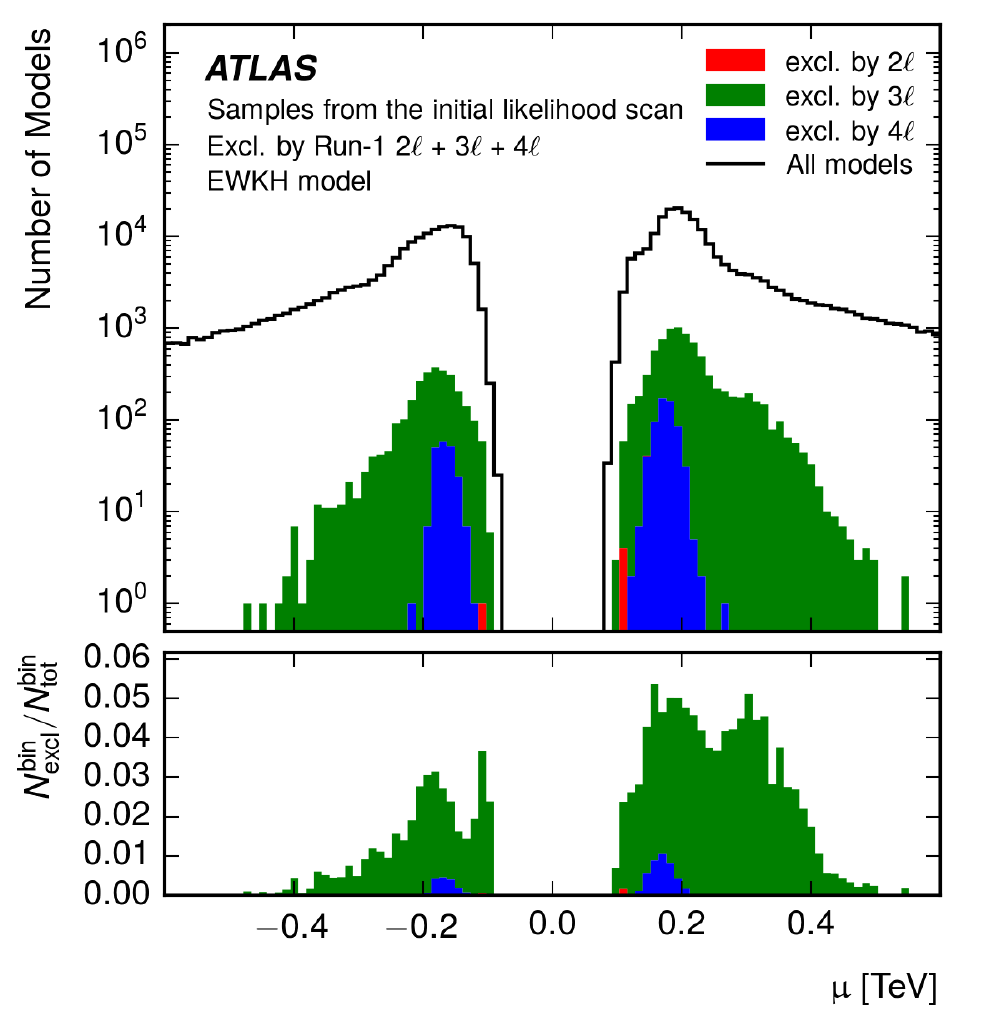}
\label{fig:resMU}
}
\subfloat[]{
\includegraphics[width=0.35\linewidth,page=1]{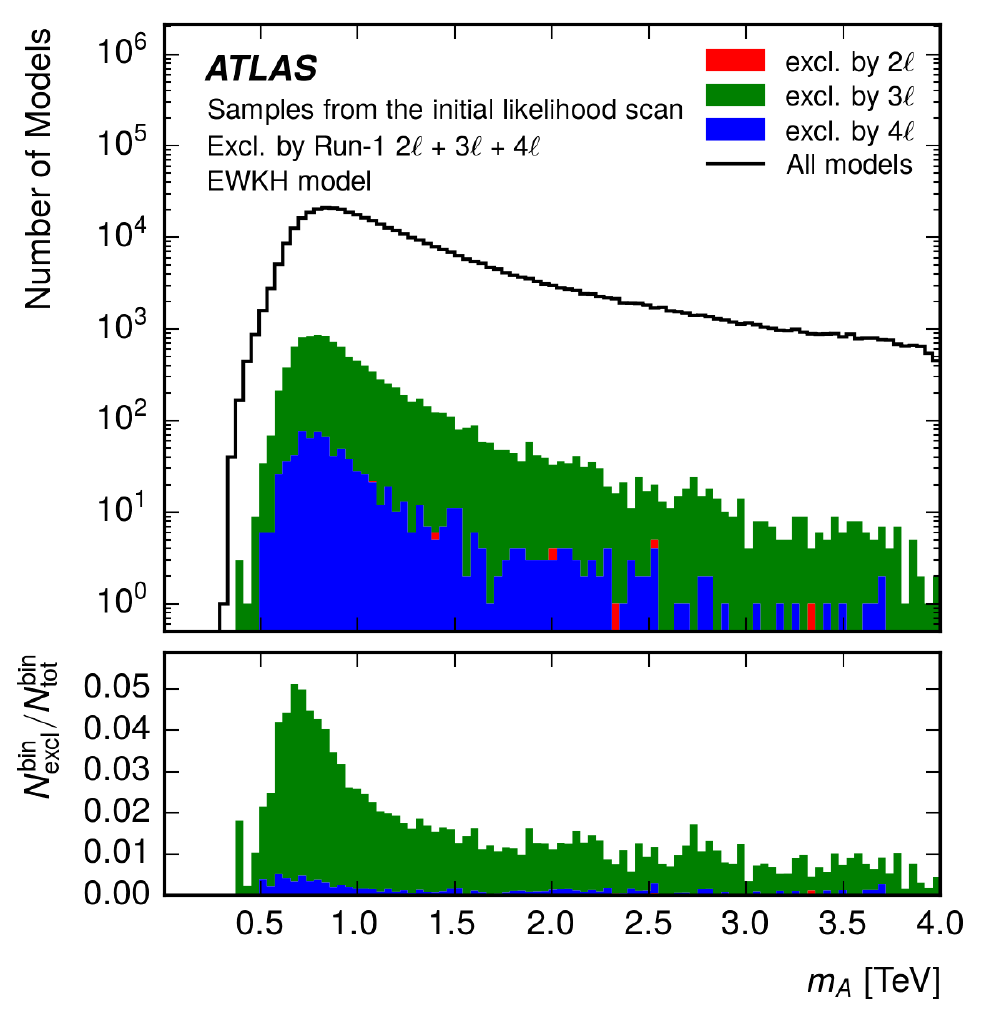}
\label{fig:resMA}
}\\
\subfloat[]{
\includegraphics[width=0.35\linewidth,page=1]{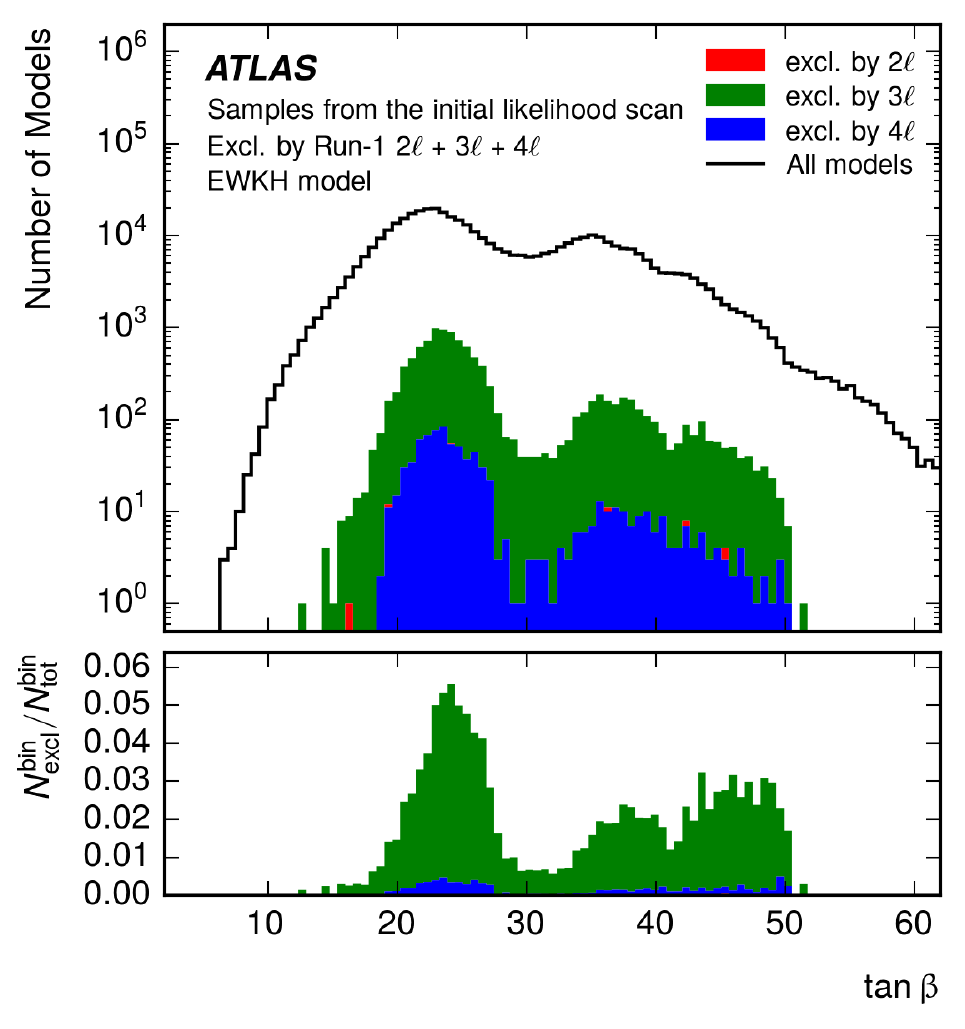}
\label{fig:resTANB}
}\\[-1ex]
\caption{
The number of models sampled by the initial likelihood scan, and the stacked bin-by-bin number of models excluded by the Run 1 ATLAS SUSY searches as a 1D function of the EWKH model parameters.
The lower part of each figure shows the fraction of models excluded by the Run 1 ATLAS SUSY searches.
The red bins indicates the fraction that is excluded by a $2\ell$ SR, the green by a $3\ell$ SR, and blue by a $4\ell$ SR.
The models considered are all within the 1D 95\% confidence interval found using the initial likelihood scan.
The plots are truncated in $|M_1|$ and $|\mu|$ to highlight the region of ATLAS electroweak SUSY sensitivity.
}
\label{fig:res1Dmodelparam}
\end{figure}

\subsection{Impact on dark matter observables}
\label{sec:res2Ddmobs}

Finally, the impact of the considered electroweak searches in several 2D parameter spaces relevant to dark matter phenomenology is shown in Figure~\ref{fig:res2Ddm}.
Figure~\ref{fig:resN1vOH2} shows the fraction of models excluded in the $\ninoone$ relic abundance versus $\ninoone$ mass plane.
The \Zboson- and $h$-funnel regions can again be clearly seen.
The exclusion power of the considered searches depends only weakly upon the relic density, which can be as small as $\sim 10^{-3}$ depending on the higgsino component of the LSP and thus the efficiency of the $s$-channel annihilation.

The region at higher LSP mass corresponds to the region where the \chinoonepm and \ninoone are close in mass (cf.\ Figure~\ref{fig:resLSPcomposition}).
Efficient coannihilation between these states (and the \ninotwo, if relevant) reduces the relic density with respect to a pure bino-like particle.
Pure higgsino-like states with $m(\ninoone) \sim 1\TeV$ and pure wino-like states with $m(\ninoone) \sim 2\TeV$ saturate the relic density.
Below these masses, mixed states can give rise to the full range of relic densities illustrated in the plot.
Finally, the $A$-funnel region can be seen in the models with $m(\ninoone) \gtrsim 200\GeV$ away from the compressed spectrum strip in Figure~\ref{fig:resC1vN1}.
In this region the $\ninoone$ is mostly bino-like.
As discussed above, the considered searches have a negligible impact on these regions.

\begin{figure}[htbp!]
\centering
\subfloat[]{
\includegraphics[width=0.45\linewidth,page=1]{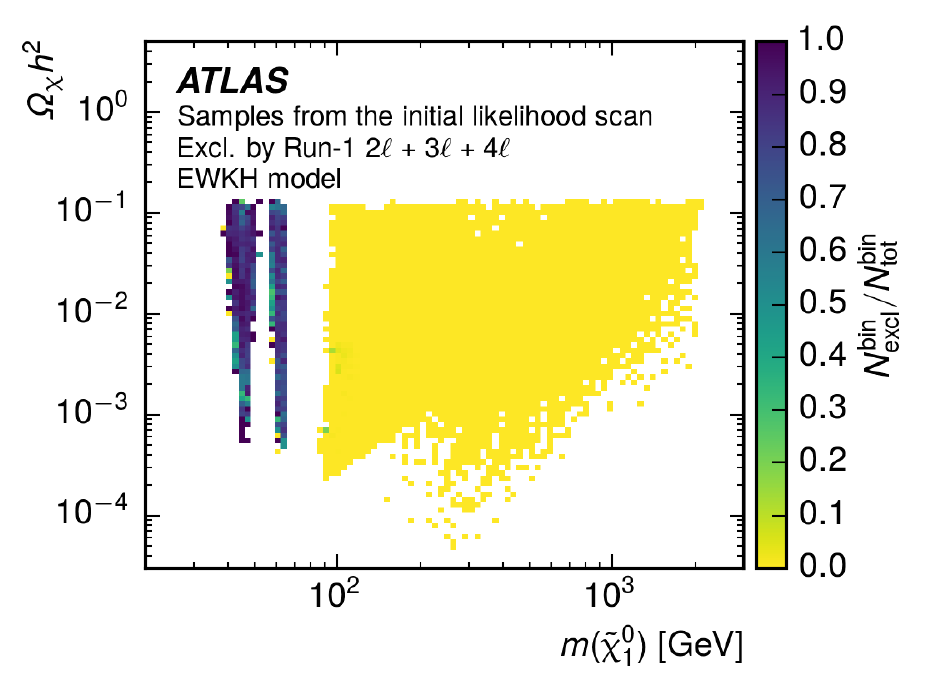}
\label{fig:resN1vOH2}
}
\subfloat[]{
\includegraphics[width=0.45\linewidth,page=1]{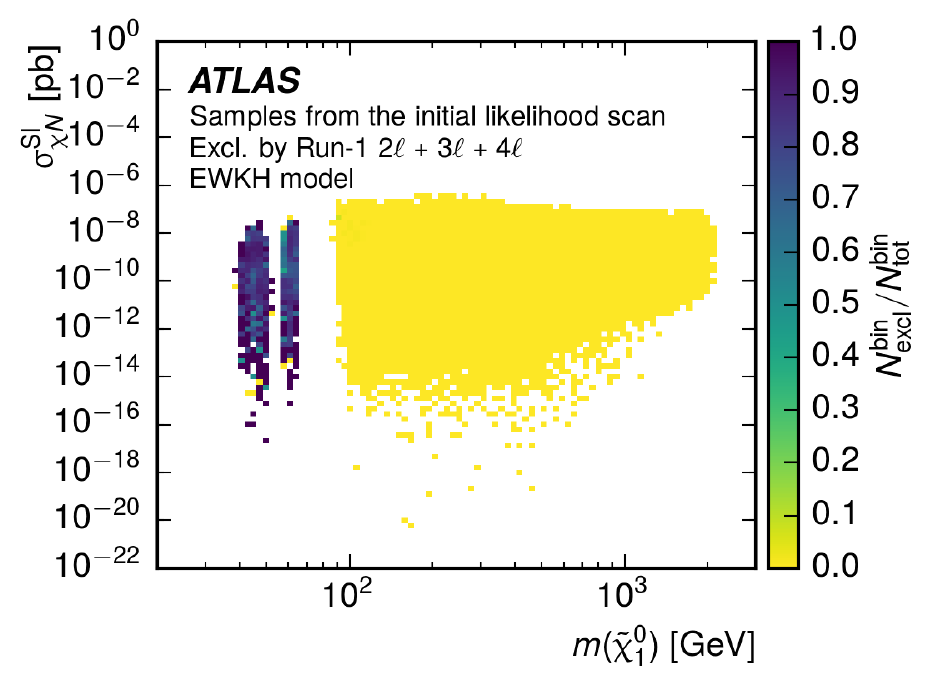}
\label{fig:resN1vSIXS}
}\\
\subfloat[]{
\includegraphics[width=0.45\linewidth,page=1]{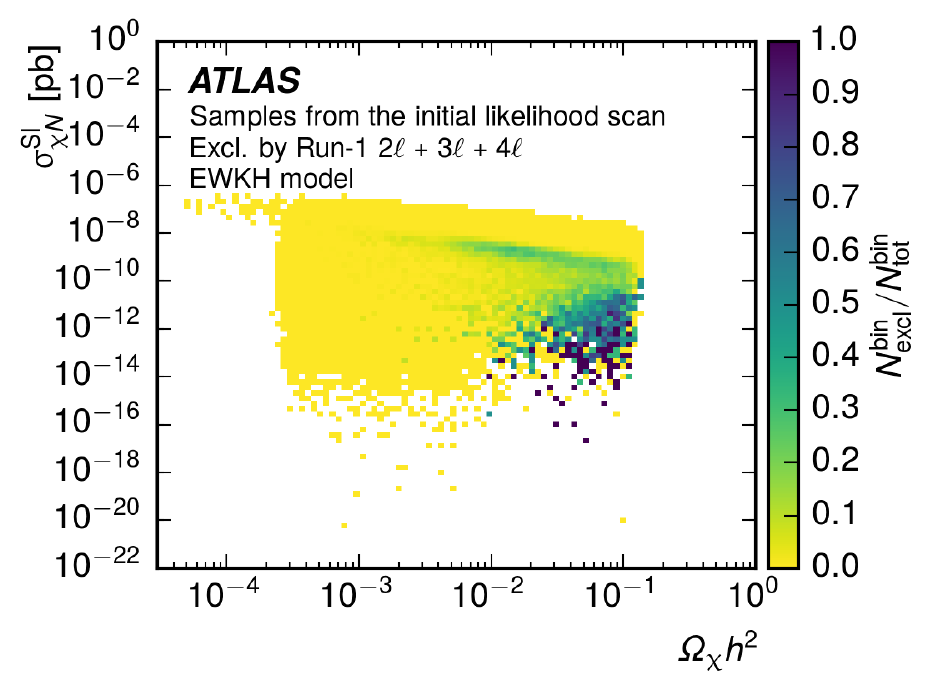}
\label{fig:resOH2vSIXS}
}\\[-1ex]
\caption{
The bin-by-bin fraction of models excluded as a 2D function of the dark matter observables.
The colour encodes the fraction of models excluded.
The models considered are all within the 2D 95\% confidence region found using the initial likelihood scan. No such models are in the white regions, and therefore the coloured bins indicate the 95\% CL contours for the initial likelihood scan.
}
\label{fig:res2Ddm}
\end{figure}

Figure~\ref{fig:resN1vSIXS} shows the SI $\ninoone$--proton scattering cross-section versus the $\ninoone$ mass.
This shows that each of the three regions with different dark matter annihilation mechanisms spans a large cross-section range.
Large values of the cross-section are not penalised in the likelihood scan because the scaling factor $\xi$ given in Equation~\eqref{eq:scaling_ansatz} reduces the predicted number of recoil events, weakening both the XENON100 and LUX constraints.
The largest cross-sections are achieved when the $\ninoone$ acquires some higgsino component, whereas cross-sections are suppressed when the $\ninoone$ has an increased bino or wino component in the low/intermediate $\ninoone$ mass regions.
Very low values of $\sigma^\text{SI}_{\chi N} \lesssim 10^{-16}\,{\rm pb}$ are rare, but occur in some models due to cancellations between the contributions from the two neutral CP-even Higgs bosons.
The SUSY searches considered here exclude a large portion of the parameter space with $m(\ninoone) \lesssim 65\GeV$, including at smaller scattering cross-sections where current and future tonne-scale underground dark matter direct-detection experiments will have less sensitivity.

Figure~\ref{fig:resOH2vSIXS} shows the ATLAS constraints in a plane of the SI $\ninoone$--proton cross-section versus the $\ninoone$ relic density.
Since this study assumes that the local $\ninoone$ density scales with the cosmological abundance, the XENON100 and LUX limits are shifted towards larger SI cross-section values for models with a relic density smaller than the value measured by Planck.
This translates into a negative correlation for large values of the SI scattering cross-section ($\gtrsim 10^{-8}\,{\rm pb}$).

For the smallest values of $\OhLSP$ ($\sim 10^{-4}$) the most favoured region of parameter space is a narrow band stretching along the currently largest allowed SI cross-section values of about $10^{-1}\,{\rm pb}$.
In this region, the low relic density is achieved by models that sit on the $A$-funnel resonance.
The $\ninoone$ for these models is mostly bino but with a sizeable higgsino content which explains the large SI cross-section.
This large SI cross-section also puts these models within reach of future direct-detection searches.
For larger relic densities, the SI cross-section is small as long as the higgsino admixture is small, but, in the case where the $\ninoone$ becomes higgsino-like (wino-like), annihilation is still efficient provided the higgsino-like (wino-like) $\ninoone$ has a mass $m(\ninoone) \lesssim 1 (2)\TeV$.
In this scenario the relic abundance is low, corresponding to the region $10^{-4} \lesssim \OhLSP \lesssim 10^{-1}$.
Finally, when the $\ninoone$ is either bino-like (with a mass of $\sim 50\GeV$ or a few hundred $\GeV$), wino-like (with a mass of about $2\TeV$)  or higgsino-like (with a mass of about $1\TeV$) then the relic density matches the measurement from the Planck Collaboration.
In these cases the SI cross-section reaches lower values because of the greater purity of the $\ninoone$.

The impact of the electroweak SUSY searches is stronger for the region of the parameter space where $10^{-2} \lesssim \OhLSP \lesssim 10^{-1}$ and the SI cross-section is low.
There is also a mild impact for larger SI cross-sections ($10^{-10}$--$10^{-8}\,{\rm pb}$) when the relic density extends down to $\OhLSP \sim 10^{-3}$.

Taken together, these results show that direct searches for the electroweak production of SUSY particles with the ATLAS experiment have a significant impact on the phenomenologically relevant \Zboson- and $h$-funnel regions in the pMSSM, without relying on the production of squarks and gluinos.
The exclusions weaken if $m(\chinoonepm) > 300\GeV$, which motivates further study with Run-2 data collected at $\sqrt{s} = 13\TeV$.
The considered searches have limited sensitivity in regions of the parameter space that favour \ninoone--\chinoonepm coannihilation and the $A$-funnel, although parts of these regions could be explored using other search channels not considered here.
The impact of the considered searches is complementary to other constraints, and they probe regions of the parameter space that are difficult to reach with direct-detection dark matter experiments.

\section{Conclusions}
\label{sec:Conclusions}
The ATLAS Collaboration performed a set of dedicated searches for electroweak SUSY particle production during the first run of the LHC, using $pp$ collisions with centre-of-mass energy of $8\TeV$ and an integrated luminosity of $20.3\,\ifb$.
In this work these searches are interpreted in a five-dimensional realisation of the pMSSM called EWKH.
This effective model parameterises the relevant dark matter phenomenology  and defines the Higgs sector at tree level of the whole pMSSM.

The parameter space of the theory was initially sampled using a likelihood-driven method.
The combined likelihood contains terms for previous collider searches, electroweak precision measurements, flavour physics results, the dark matter relic density and direct dark matter searches, as well as the Higgs boson mass.
The dimensionality of the initial likelihood scan was reduced to one or two parameters by maximising the likelihood function over the remaining parameters. 
This produced $472\,933$ models within the 2D $95\%$ confidence-level region.

Constraints from ATLAS searches for electroweak SUSY particle production in events with two, three and four charged leptons were then applied by taking the \cls\ value of the signal region with the best expected sensitivity for each model. Models with $\cls < 0.05$ were considered to be excluded at 95\% confidence level.
Due to the number of models involved, a new method for estimating the \cls\ values of different signal regions was developed, which uses only generator-level information in addition to a calibration sample of 500 models that were passed through the full ATLAS detector simulation.
The fraction of models excluded as a function of model parameters and masses was then studied.

The dark matter relic density measurement from the Planck Collaboration allows only four regions of parameter space. These correspond to three mechanisms that achieve a high enough dark matter annihilation cross-section.
These regions are: the \Zboson-funnel with $m(\ninoone) \approx 45\GeV$; the $h$-funnel with $m(\ninoone) \approx 60\GeV$; the coannihilation region with $m(\ninoone) \approx m(\chinoonepm)$ which extends up to $m(\ninoone) \approx 2\TeV$; and the $A$-funnel with $0.2  \lesssim  m(\ninoone) \lesssim 2\TeV$.
The considered searches exclude $86\%$ of the models in the \Zboson- and $h$-funnel regions ($m(\ninoone) < 65\GeV$) while having negligible sensitivity to the coannihilation and $A$-funnel regions.
The mass spectrum in the coannihilation region is, by definition, compressed, and any leptons produced in the decays of these particles are too soft for the considered searches.
It is possible that an existing search, not considered here, for events with a disappearing track would be sensitive to the portion of this region where the mass splitting is $\lesssim 200\MeV$, leading to a metastable chargino that would decay in the detector.
In addition, it has been shown that ATLAS searches for squark and gluino production can constrain this region of parameter space, if strongly interacting sparticles are accessible at the LHC.
Similarly, in the absence of strong production the $A$-funnel region is inaccessible because the electroweakinos are too heavy to be detectable with current data.
In all, values of $|M_1|$ below $100\GeV$ are strongly constrained by the considered searches, while the constraints on $M_2$, $\mu$ and the other parameters are less stringent.

Accelerator searches are found to be complementary to direct-detection constraints.
In particular, a region of the parameter space with $m(\ninoone) \lesssim 65\GeV$ and small values of the spin-independent interaction cross-section are probed by the Run-1 LHC searches.
The different regions of sensitivity demonstrate clearly the importance of these complementary experimental techniques in dark matter searches.

\FloatBarrier

\section*{Acknowledgements}

We thank CERN for the very successful operation of the LHC, as well as the
support staff from our institutions without whom ATLAS could not be
operated efficiently.

We acknowledge the support of ANPCyT, Argentina; YerPhI, Armenia; ARC, Australia; BMWFW and FWF, Austria; ANAS, Azerbaijan; SSTC, Belarus; CNPq and FAPESP, Brazil; NSERC, NRC and CFI, Canada; CERN; CONICYT, Chile; CAS, MOST and NSFC, China; COLCIENCIAS, Colombia; MSMT CR, MPO CR and VSC CR, Czech Republic; DNRF and DNSRC, Denmark; IN2P3-CNRS, CEA-DSM/IRFU, France; GNSF, Georgia; BMBF, HGF, and MPG, Germany; GSRT, Greece; RGC, Hong Kong SAR, China; ISF, I-CORE and Benoziyo Center, Israel; INFN, Italy; MEXT and JSPS, Japan; CNRST, Morocco; FOM and NWO, Netherlands; RCN, Norway; MNiSW and NCN, Poland; FCT, Portugal; MNE/IFA, Romania; MES of Russia and NRC KI, Russian Federation; JINR; MESTD, Serbia; MSSR, Slovakia; ARRS and MIZ\v{S}, Slovenia; DST/NRF, South Africa; MINECO, Spain; SRC and Wallenberg Foundation, Sweden; SERI, SNSF and Cantons of Bern and Geneva, Switzerland; MOST, Taiwan; TAEK, Turkey; STFC, United Kingdom; DOE and NSF, United States of America. In addition, individual groups and members have received support from BCKDF, the Canada Council, CANARIE, CRC, Compute Canada, FQRNT, and the Ontario Innovation Trust, Canada; EPLANET, ERC, FP7, Horizon 2020 and Marie Sk{\l}odowska-Curie Actions, European Union; Investissements d'Avenir Labex and Idex, ANR, R{\'e}gion Auvergne and Fondation Partager le Savoir, France; DFG and AvH Foundation, Germany; Herakleitos, Thales and Aristeia programmes co-financed by EU-ESF and the Greek NSRF; BSF, GIF and Minerva, Israel; BRF, Norway; Generalitat de Catalunya, Generalitat Valenciana, Spain; the Royal Society and Leverhulme Trust, United Kingdom.

The crucial computing support from all WLCG partners is acknowledged gratefully, in particular from CERN, the ATLAS Tier-1 facilities at TRIUMF (Canada), NDGF (Denmark, Norway, Sweden), CC-IN2P3 (France), KIT/GridKA (Germany), INFN-CNAF (Italy), NL-T1 (Netherlands), PIC (Spain), ASGC (Taiwan), RAL (UK) and BNL (USA), the Tier-2 facilities worldwide and large non-WLCG resource providers. Major contributors of computing resources are listed in Ref.~\cite{ATL-GEN-PUB-2016-002}.

\printbibliography
\newpage
\begin{flushleft}
{\Large The ATLAS Collaboration}

\bigskip

M.~Aaboud$^\textrm{\scriptsize 136d}$,
G.~Aad$^\textrm{\scriptsize 87}$,
B.~Abbott$^\textrm{\scriptsize 114}$,
J.~Abdallah$^\textrm{\scriptsize 8}$,
O.~Abdinov$^\textrm{\scriptsize 12}$,
B.~Abeloos$^\textrm{\scriptsize 118}$,
R.~Aben$^\textrm{\scriptsize 108}$,
O.S.~AbouZeid$^\textrm{\scriptsize 138}$,
N.L.~Abraham$^\textrm{\scriptsize 152}$,
H.~Abramowicz$^\textrm{\scriptsize 156}$
% ...
\bigskip
\\
% ...
$^{8}$ Department of Physics, The University of Texas at Arlington, Arlington TX, United States of America\\
% ...
$^{12}$ Institute of Physics, Azerbaijan Academy of Sciences, Baku, Azerbaijan\\
% ...
$^{87}$ CPPM, Aix-Marseille Universit{\'e} and CNRS/IN2P3, Marseille, France\\
% ...
$^{108}$ Nikhef National Institute for Subatomic Physics and University of Amsterdam, Amsterdam, Netherlands\\
% ...
$^{114}$ Homer L. Dodge Department of Physics and Astronomy, University of Oklahoma, Norman OK, United States of America\\
% ...
$^{118}$ LAL, Univ. Paris-Sud, CNRS/IN2P3, Universit{\'e} Paris-Saclay, Orsay, France\\
% ...
$^{136}$ $^{(a)}$ Facult{\'e} des Sciences Ain Chock, R{\'e}seau Universitaire de Physique des Hautes Energies - Universit{\'e} Hassan II, Casablanca; $^{(b)}$ Centre National de l'Energie des Sciences Techniques Nucleaires, Rabat; $^{(c)}$ Facult{\'e} des Sciences Semlalia, Universit{\'e} Cadi Ayyad, LPHEA-Marrakech; $^{(d)}$ Facult{\'e} des Sciences, Universit{\'e} Mohamed Premier and LPTPM, Oujda; $^{(e)}$ Facult{\'e} des sciences, Universit{\'e} Mohammed V, Rabat, Morocco\\
% ...
$^{138}$ Santa Cruz Institute for Particle Physics, University of California Santa Cruz, Santa Cruz CA, United States of America\\
% ...
$^{152}$ Department of Physics and Astronomy, University of Sussex, Brighton, United Kingdom\\
% ...
$^{156}$ Raymond and Beverly Sackler School of Physics and Astronomy, Tel Aviv University, Tel Aviv, Israel\\
% ...
$^{a}$ Also at Department of Physics, King's College London, London, United Kingdom\\
$^{b}$ Also at Institute of Physics, Azerbaijan Academy of Sciences, Baku, Azerbaijan\\
$^{c}$ Also at Novosibirsk State University, Novosibirsk, Russia\\
$^{d}$ Also at TRIUMF, Vancouver BC, Canada\\
$^{e}$ Also at Department of Physics {\&} Astronomy, University of Louisville, Louisville, KY, United States of America\\
% ...
\end{flushleft}


\end{document}